\documentclass{iopart}
\usepackage{iopams}

\begin{document}
\title{Einstein-Schr\"{o}dinger theory using Newman-Penrose tetrad formalism}
\author{J. A. Shifflett}
\address{
Department of Physics,
Washington University, St.~Louis, Missouri 63130
}
\begin{abstract}
The Einstein-Schr\"{o}dinger theory is modified to include a large cosmological
constant caused by zero-point fluctuations. This ``extrinsic'' cosmological
constant which multiplies the symmetric metric is assumed to be nearly cancelled
by Schr\"{o}dinger's ``bare'' cosmological constant which multiplies the
nonsymmetric fundamental tensor, such that the total cosmological constant is
consistent with measurement. This modified Einstein-Schr\"{o}dinger theory is
expressed in Newman-Penrose form, and tetrad methods are used to confirm that
it closely approximates ordinary general relativity and electromagnetism.
A solution for the connections in terms of the fundamental tensor is derived
in the tetrad frame. The tetrad form of an exact electric monopole solution is
shown to approximate the Reissner-Nordstr\"{o}m solution and to be of Petrov type-D.
\end{abstract}
\newcommand{\K}{N}
\newcommand{\NPD}{D}
\newcommand{\NPDEL}{\Delta}
\newcommand{\NPdel}{\delta}
\newcommand{\NPdels}{\delta^*}
\newcommand{\dual}{\vartheta}
\newcommand{\tGam}{{^\circ\Gamma}}
\newcommand{\tPi}{{^\circ\Pi}}
\newcommand{\tR}{{^\circ\!R}}
\newcommand{\tG}{{^\circ\!G}}
\newcommand{\N}{N}
\newcommand{\Nbar}{W}
\newcommand{\J}{j}
\newcommand{\rmg}{\sqrt{-g}}
\newcommand{\rmN}{\sqrt{-\N}}
\newcommand{\rmgt}{\sqrt{-g_\diamond}}
\newcommand{\rmNt}{\sqrt{-\N_\diamond}}
\newcommand{\Q}{Q}
\newcommand{\ff}{\,\ell}
\newcommand{\Aphi}{A}
\newcommand{\OPMU}{(1\!\pm\!\tanht)}
\newcommand{\OMPU}{(1\!\mp\!\tanht)}
\newcommand{\OPMIV}{(1\!\pm\!i\tant)}
\newcommand{\OMPIV}{(1\!\mp\!i\tant)}
\newcommand{\OPU}{(1\!+\!\tanht)}
\newcommand{\OMU}{(1\!-\!\tanht)}
\newcommand{\OPIV}{(1\!+\!i\tant)}
\newcommand{\OMIV}{(1\!-\!i\tant)}
\newcommand{\HD}{\varpi}
\newcommand{\uacute}{\acute{u}}
\newcommand{\tanht}{\check u}
\newcommand{\cosht}{\check c}
\newcommand{\sinht}{\check s}
\newcommand{\Zht}{\check z}
\newcommand{\tant}{\grave u}
\newcommand{\cost}{\grave c}
\newcommand{\sint}{\grave s}
\newcommand{\Zt}{\grave z}
\newcommand{\dete}{{\bf e}}
\newcommand{\FO}{O}
\newcommand{\mum}{\mu}
% Einstein Maxwell spacetimes (04.40.Nr)
% cosmological constant (98.80.Es)
% dark matter (95.35.+d)
% exact solutions of general relativity (04.20.Jb)
% unified field theories and models (12.10.-g)
% relativity and gravitation (95.30.Sf) - probably covered by (04.40.Nr)
% alternative theories of gravity (04.50.+h) - it isn't about gravity
% No more than 4 of these are supposed to be used
% PACS codes
%\keywords{Einstein-Schrodinger Theory, Hermitian Theory of Relativity,
%Schrodinger Affine Field Theory, Einstein-Straus Theory,
%Cosmological Constant, Zero-Point Fluctuations}
%Use showkeys class option
\ead{shifflet@hbar.wustl.edu}
\section{\label{Introduction}Introduction}

The Einstein-Schr\"{o}dinger theory without a cosmological constant
was originally proposed by Einstein and Straus in
1946\cite{EinsteinStraus,Einstein3,EinsteinBianchi,EinsteinKaufman,EinsteinMOR}.
%In this form it is also known as Einstein's Unified Field Theory, or the
%Einstein-Straus Theory, or the Hermitian Theory of Relativity, or the
%Nonsymmetric Gravitational Theory, or by Einstein's
%own name, the Generalized Theory of Gravitation.
Schr\"{o}dinger made an important contribution to the theory by
generalizing it to include a cosmological constant, and by showing
that the theory can be derived from a very simple Lagrangian density
if this cosmological constant is assumed to be
non-zero\cite{SchrodingerI,SchrodingerII,SchrodingerIII,SchrodingerSTS}.
This more general theory is usually called Schr\"{o}dinger's Affine Field Theory
or the Einstein-Schr\"{o}dinger Theory. This theory is a generalization of
ordinary general relativity which allows a non-symmetric fundamental tensor and
connection. Einstein and Schr\"{o}dinger suspected that the antisymmetric part
of the fundamental tensor might contain the electromagnetic field, but despite
much effort this has never been demonstrated.

Recently we have shown\cite{Shifflett,Shifflett3} that a well motivated
modification of the Einstein-Schr\"{o}dinger theory does indeed closely
approximate ordinary general relativity and electromagnetism, the modification
being the addition of a cosmological constant caused by zero-point fluctuations.
It is reasonable to assume that the Einstein-Schr\"{o}dinger
theory must eventually be quantized to accurately predict reality, and this
cosmological constant can be viewed as a kind of zeroth order
quantization effect\cite{Sahni,Peskin,Zeldovich}.
This ``extrinsic'' cosmological constant which multiplies the symmetric metric
is assumed to be nearly cancelled by Schr\"{o}dinger's ``bare'' cosmological
constant which multiplies the nonsymmetric fundamental tensor, resulting in a
total cosmological constant which is consistent with measurement. The fact that
these two cosmological constants multiply different fields has the effect of
creating a Lorentz force, and also fixes some other problems with the original
theory. The fine-tuning of cosmological constants is less objectionable when one
considers that it is similar to renormalization methods which are commonplace in
quantum field theory. For example, to account for self energy, ``bare'' particle
masses become large (infinite if the cutoff wavenumber goes to infinity), but
the total ``physical'' mass remains small. In a similar manner in the present
theory, Schr\"{o}dinger's ``bare'' cosmological constant becomes large, but the
total ``physical'' cosmological constant remains small. This can
be viewed as a kind of energy density renormalization of the original
Einstein-Schr\"{o}dinger theory to account for zero-point fluctuations, and with
this quantization effect included, the theory closely
approximates ordinary general relativity and
electromagnetism.

As in \cite{Shifflett,Shifflett3} and in papers by several other
authors\cite{Hely,Treder57,JohnsonI,Antoci3},
the metric is defined as,
\begin{eqnarray}
\label{gdef}
\fl g^{\rho\tau}\!=\!\frac{\rmN}{\rmg} \K^{\dashv(\rho\tau)}
,~~\rmg\!=\![-det(\rmg g^{\mu\nu})]^{1/(n-2)}
\!=\![-det(\rmN \K^{\dashv(\mu\nu)})]^{1/(n-2)}.
%g^{\rho\tau}=\frac{\rmN}{\rmg}\K^{\dashv(\rho\tau)}.
\end{eqnarray}
Here $N_{\mu\nu}$ is the fundamental tensor, $\N\!\!=\!det(\N_{\mu\nu})$,
$g\!\!=\!det(g_{\mu\nu})$, ``n'' is the dimension,
and $\K^{\dashv\rho\tau}$ is the inverse of
$N_{\tau\mu}$ so that $\K^{\dashv\rho\tau}N_{\tau\mu}\!=\!\delta^\rho_\mu$.
%so that $\K^{\dashv\mu\rho}N_{\sigma\mu}\!=\!\delta^\rho_\sigma$
%$\N_{\sigma\mu}$ so that $\N^{\sigma\nu}\N_{\sigma\mu}\!=\!\delta^\nu_\mu$.
When $N_{\mu\nu}$ is symmetric, the equation
gives $g_{\mu\nu}\!=\!N_{\mu\nu}$ as desired.
%The metric (\ref{gdef})
%results in the motion of neutral particles being unaffected by
%$\N_{[\sigma\mu]}$\cite{JohnsonI}, and has several other desirable
%properties\cite{Hely,Treder57,Antoci3,Shifflett3}.
In this paper, raising and lowering of indices is always done using
(\ref{gdef}), and covariant derivative ``;'' is done using the Christoffel
connection $\Gamma^\alpha_{\sigma\mu}$ formed from (\ref{gdef}).

The field equations of the Einstein-Schr\"{o}dinger theory,
including an extrinsic cosmological term\cite{Shifflett}, energy-momentum
tensor\cite{Treder57}, and charge currents\cite{Borchsenius,Antoci3} are,
\begin{eqnarray}
\label{JSsymmetric}
\fl \tR_{(\sigma\mu)}
+\Lambda_b \N_{(\sigma\mu)}+\Lambda_e g_{\sigma\mu}
=\frac{8\pi G}{c^4}\!\left(T_{\sigma\mu}
-\frac{1}{(n\!-\!2)}g_{\sigma\mu}T^\alpha_\alpha\right),\\
\label{JScurl}
\fl \tR_{[\sigma\mu,\nu]}+\Lambda_b \N_{[\sigma\mu,\nu]}=0,\\
\label{JSconnection}
\fl \N_{\sigma\mu,\beta}\!-\!\tGam^\alpha_{\sigma\beta}\N_{\alpha\mu}
\!-\!\tGam^\alpha_{\beta\mu}\N_{\sigma\alpha}
=-\frac{8\pi}{c(n\!-\!1)}\frac{\rmg}{\rmN}\!\left(
\N_{\sigma[\alpha}\N_{\beta]\mu}
\!+\frac{1}{(n\!-\!2)} \N_{[\alpha\beta]}\N_{\sigma\mu}\!\right)\!\J^\alpha,\\
\label{JScontractionsymmetric}
\fl \tGam^\alpha_{\beta\alpha}=\tGam^\alpha_{\alpha\beta}.
\end{eqnarray}
Here $\tR_{\sigma\mu}\!=\!R_{\sigma\mu}(\tGam)$ is the Ricci tensor,
$\tGam^\alpha_{\sigma\mu}$ is the non-symmetric connection, $T_{\sigma\mu}$
is an energy-momentum tensor with no electromagnetic component, and $\J^\alpha$
is a charge current.
%, and ``n'' is the dimension.
Like Schr\"{o}dinger,
we include a bare cosmological constant $\Lambda_b$ because
this allows a very simple derivation of the theory\cite{SchrodingerI,Shifflett}.
%We also include a large extrinsic cosmological constant $\Lambda_e$,
%perhaps caused by zero-point fluctuations\cite{Shifflett}.
The total $\Lambda$ is then
\begin{eqnarray}
&&\Lambda=\Lambda_b+\Lambda_e.
\end{eqnarray}

From a theorem of tensor calculus\cite{Adler}, (\ref{JScurl}) implies that
$\tR_{[\sigma\mu]}+\Lambda_b \N_{[\sigma\mu]}$ is a curl,
so (\ref{JScurl}) can be written in the completely equivalent form
\begin{eqnarray}
\label{paraantisymmetric}
&&\tR_{[\sigma\mu]}
+\tGam^\alpha_{\alpha[\sigma,\mu]}
+2\Lambda_b\Aphi_{[\sigma,\mu]}
+\Lambda_b \N_{[\sigma\mu]}=0,
\end{eqnarray}
and (\ref{JSsymmetric},\ref{JScurl}) can be
combined together in the completely equivalent form,
\begin{eqnarray}
\label{para}
\fl~~ \tR_{\sigma\mu}
+\tGam^\alpha_{\alpha[\sigma,\mu]}
+2\Lambda_b\Aphi_{[\sigma,\mu]}
+\Lambda_b \N_{\sigma\mu}+\Lambda_e g_{\sigma\mu}
=\frac{8\pi G}{c^4}\!\left(T_{\sigma\mu}
-\frac{1}{(n\!-\!2)}g_{\sigma\mu}T^\alpha_\alpha\right).
\end{eqnarray}
The tensor term $\tGam^\alpha_{\alpha[\sigma,\mu]}$ is needed
to retain Hermitian
symmetry\cite{EinsteinStraus,Borchsenius,Antoci3,Shifflett3},
and $\tR_{\sigma\mu}\!+\tGam^\alpha_{\alpha[\sigma,\mu]}$
is sometimes called the Hermitianized Ricci tensor.
%because it has the property
%$\tPi_{\sigma\mu}(\tGam^T)=\tPi_{\mu\sigma}(\tGam)$.
Contracting (\ref{JSconnection}) with $\K^{\dashv\mu\sigma}\!/2$ shows that
this extra term vanishes when charge currents are absent,
\begin{eqnarray}
\label{selftransplant}
\fl\tGam^\alpha_{\alpha\beta}
\!-\!\frac{8\pi}{c(n\!-\!1)(n\!-\!2)}\frac{\rmg}{\rmN}
\J^\rho \N_{[\rho\beta]}
\!=\!\frac{1}{2}\K^{\dashv\mu\sigma}\N_{\sigma\mu,\beta}
\!=\!\frac{1}{2\N}\frac{\partial\N}{\partial \N_{\sigma\mu}}\N_{\sigma\mu,\beta}
%\!=\!\frac{\N_{\!,\beta}}{2\N}
\!=\!\frac{(\rmN\,)_{,\beta}}{\rmN},\\
\label{RisB}
\fl\tGam^\alpha_{\alpha[\sigma,\mu]}
-\frac{8\pi}{c(n\!-\!1)(n\!-\!2)}
\left(\!\frac{\rmg}{\rmN}\J^\rho \N_{[\rho[\sigma]}\!\right)\!\!{_{,\mu]}}
=(ln\rmN)_{,[\sigma,\mu]}=0.
\end{eqnarray}

Multiplying (\ref{JSconnection}) by $-\K^{\dashv\rho\sigma}\K^{\dashv\mu\tau}$
and using (\ref{selftransplant}) gives
\begin{eqnarray}
\label{contravariant}
\fl {\K^{\dashv\rho\tau}}{_{,\beta}}
\!+\!\tGam^\tau_{\nu\beta}\K^{\dashv\rho\nu}
\!+\!\tGam^\rho_{\beta\nu}\K^{\dashv\nu\tau}
=\frac{8\pi}{c(n\!-\!1)}\frac{\rmg}{\rmN}\!\left(
\J^{[\rho}\delta^{\tau]}_\beta
+\frac{1}{(n\!-\!2)}\J^\alpha\N_{[\alpha\beta]}\K^{\dashv\rho\tau}\!\right)\!,\\
\label{contravariantdensity}
\fl(\!\rmN \K^{\dashv\rho\tau})_{\!,\,\beta}
\!+\!\tGam^\tau_{\nu\beta}\rmN \K^{\dashv\rho\nu}
\!+\!\tGam^\rho_{\beta\nu}\rmN \K^{\dashv\nu\tau}
\!-\!\tGam^\alpha_{\beta\alpha}\rmN \K^{\dashv\rho\tau}\nonumber\\
\fl~~~~~~~~~~~~~~~~~~~~~~~~~~~~~~~~~~~~~~~~~~~~~~~~~~~~~~~~~~~~~~~~~
\!=\!\frac{8\pi}{c(n\!-\!1)\!}\rmg \J^{[\rho}\delta^{\tau]}_\beta.
\end{eqnarray}
Taking the antisymmetric part of (\ref{contravariantdensity})
and contracting gives Ampere's law
\begin{eqnarray}
\label{Ampere}
f^{\tau\rho}{_{;\tau}}
&=&\frac{(\rmg f^{\tau\rho}){_{,\tau}}}{\rmg}
=\frac{(\rmN\,\K^{\dashv[\rho\tau]}){_{,\tau}}}{\rmg}
=\frac{4\pi}{c}\J^\rho
\end{eqnarray}
and the continuity equation
\begin{eqnarray}
\label{continuity}
\J^\rho{_{\!;\rho}}&=&(c/4\pi)f^{\tau\rho}{_{;[\tau;\rho]}}=0,
\end{eqnarray}
where we define
\begin{eqnarray}
\label{fdef}
f^{\rho\tau}=-\frac{\rmN}{\rmg}\K^{\dashv[\rho\tau]}.
\end{eqnarray}

The Einstein equations are obtained by combining (\ref{JSsymmetric})
with its contraction,
\begin{eqnarray}
\label{Einstein}
\fl~~~~~~~~~~\frac{8\pi G}{c^4}T_{\sigma\mu}
=\tG_{\sigma\mu}
+\Lambda_b\!\left(\N_{(\sigma\mu)}
\!-\!\frac{1}{2}g_{\sigma\mu}\N^\rho_\rho\right)
\!+\!\Lambda_e\!\left(1-\frac{n}{2}\right)g_{\sigma\mu},
\end{eqnarray}
where we define
\begin{eqnarray}
\label{genEinstein}
\tG_{\sigma\mu}
=\tR_{(\sigma\mu)}
-\frac{1}{2}g_{\sigma\mu}\tR^\rho_\rho.
\end{eqnarray}
Using (\ref{JSconnection},\ref{JScontractionsymmetric},\ref{gdef},\ref{fdef}),
a generalized contracted Bianchi identity with charge currents can be
derived for this theory\cite{Antoci3,Shifflett3},
\begin{eqnarray}
\label{contractedBianchi}
\!\!\!\tG^\nu_{\sigma;\,\nu}
=\frac{3}{2}f^{\nu\rho}\tR_{[\nu\rho,\sigma]}
+\frac{4\pi}{c}\J^\nu(\tR_{[\nu\sigma]}
+\tGam^\alpha_{\alpha[\nu,\sigma]}).
\end{eqnarray}
Using only the definitions (\ref{gdef},\ref{fdef}),
another useful identity is derived in \cite{Shifflett3},
\begin{eqnarray}
\label{usefulidentity}
\left(N^{(\mu}{_{\sigma)}}
\!-\!\frac{1}{2}\delta^\mu_\sigma N^\rho_\rho\right)\!{_{;\,\mu}}
-\frac{3}{2}f^{\nu\rho}N_{[\nu\rho,\sigma]}
=f^{\nu\rho}{_{;\nu}}N_{[\rho\sigma]}.
\end{eqnarray}
Using (\ref{contractedBianchi},\ref{JScurl},
\ref{usefulidentity},\ref{Ampere},\ref{paraantisymmetric}),
the divergence of the Einstein equations (\ref{Einstein})
gives the ordinary Lorentz force equation of
general relativity and electromagnetism\cite{Shifflett3}
\begin{eqnarray}
\label{divergence}
\fl~~~~~\frac{8\pi G}{c^4}T^\nu_{\sigma;\,\nu}
=\frac{4\pi}{c}\J^\nu(\tR_{[\nu\sigma]}
+\tGam^\alpha_{\alpha[\nu,\sigma]})
+\Lambda_b\frac{4\pi}{c}\J^\nu \N_{[\nu\sigma]}
=\frac{8\pi}{c}\Lambda_b\J^\nu A_{[\sigma,\nu]}.
\end{eqnarray}
Assuming $T_{\alpha\nu}\!=\!\mum c^2 u_\alpha u_\nu$, with
$u^\alpha\!=\!dx^\alpha/ds\!=\!\J^\alpha m/ec\mu$ and
$(\mum u^\nu)_{;\nu}\!=\!0$ from (\ref{continuity}), equation
(\ref{divergence}) is just the Lorentz force coupled to Newton's 2nd law,
\begin{eqnarray}
\label{Lorentz}
\fl~~~~~\frac{c^3}{G}\Lambda_b\J^\nu A_{[\sigma,\nu]}
=T^\nu_{\sigma;\,\nu}
=(\mum c^2 u^\nu u_{\sigma})_{;\,\nu}
=\mum c^2 u^\nu u_{\sigma;\,\nu}
=\frac{\mum}{m}\left(m\frac{c\,dx_\sigma}{ds}\right){_{\!;\nu}}
\frac{c\,dx^\nu}{ds}.
\end{eqnarray}
Here the conversion to cgs units is,
\begin{eqnarray}
\label{cgs}
\fl
f_{\sigma\mu}
\!=\!\sqrt{\frac{\lower2pt\hbox{$-2G$}}{c^4\Lambda_b}}f_{\sigma\mu}^{\,(cgs)}
,~
A_{\sigma}
\!=\!\sqrt{\frac{\lower2pt\hbox{$-2G$}}{c^4\Lambda_b}}A_{\sigma}^{(cgs)}
,~
\J_{\sigma}
\!=\!\sqrt{\frac{\lower2pt\hbox{$-2G$}}{c^4\Lambda_b}}\,\J_{\sigma}^{\,(cgs)}
,~
Q
\!=\!\sqrt{\frac{\lower2pt\hbox{$-2G$}}{c^4\Lambda_b}}\,Q^{(cgs)}.
%~~~~~{\rm (and~likewise~for~A_\sigma~and~\J^\mu}).
\end{eqnarray}

Eq.~(\ref{JSconnection}) describes an implicit algebraic dependence
of $\tGam^\alpha_{\sigma\beta}$ on $\N_{\sigma\mu}$
and $\N_{\sigma\mu,\beta}$. The solution for
$\tGam^\alpha_{\sigma\beta}(\N_{\mu\nu})$ yields the Christoffel
connection for the symmetric case,
\begin{eqnarray}
\label{Christoffel}
\Gamma^\alpha_{\sigma\mu}&=&\frac{1}{2}g^{\alpha\nu}(g_{\mu\nu,\sigma}
+g_{\nu\sigma,\mu}-g_{\sigma\mu,\nu}).
\end{eqnarray}
There is also a fairly simple solution\cite{Hlavaty} for the special
case $f^\sigma{_\mu}f^\mu{_\sigma}\!=det(f^{\mu}{_{\nu}})\!=0$.
The solution for $\tGam^\alpha_{\sigma\beta}(\N_{\mu\nu})$ is much more
complicated in the general non-symmetric case\cite{Tonnelat,Hlavaty}, and this
has been a big obstacle in working with the theory. Here we will derive a
solution in the Newman-Penrose tetrad frame which applies for all cases except
$f^\sigma{\!_\mu}f^\mu{\!_\sigma}\!=det(f^{\mu}{_{\nu}})\!=\!0$.
It is given as an addition $\Upsilon^\tau_{\nu\beta}$
to the Christoffel connection,
\begin{eqnarray}
\label{sum}
\tGam^\tau_{\nu\beta}=\Gamma^\tau_{\nu\beta}+\Upsilon^\tau_{\nu\beta}.
\end{eqnarray}
Extracting $\Upsilon^\tau_{\nu\beta}$ from the Ricci tensor
gives\cite{Shifflett},
\begin{eqnarray}
\label{Ricciaddition}
\fl~~~~~~~~~~~\tR_{\sigma\mu}
+\tGam^\alpha_{\alpha[\sigma,\mu]}
&=&R_{\sigma\mu}+\Upsilon^\alpha_{\sigma\mu;\alpha}
\!-\Upsilon^\alpha_{\alpha(\sigma;\mu)}
\!-\Upsilon^\nu_{\sigma\alpha}\Upsilon^\alpha_{\nu\mu}
\!+\Upsilon^\nu_{\sigma\mu}\Upsilon^\alpha_{\nu\alpha}.
\end{eqnarray}
Here $R_{\sigma\mu}=R_{\sigma\mu}(\Gamma)$ is the ordinary Ricci tensor.
Substituting (\ref{Ricciaddition}) into (\ref{para}) or into
(\ref{JSsymmetric},\ref{JScurl},\ref{paraantisymmetric},\ref{Einstein}),
and working in the tetrad frame, one then has an explicit version of the field
equations. It must be mentioned that the
solution for $\tGam^\alpha_{\sigma\beta}(\N_{\mu\nu})$ in the Newman-Penrose
tetrad frame has in a sense already been done in \cite{Hlavaty}, prior to
the development of the full Newman-Penrose formalism\cite{Newman}. However,
\cite{Hlavaty} uses a different metric definition and does not
include charge currents. This reference also uses completely non-standard
conventions and notation, and does not simplify results.

This paper is organized as follows. In \S\ref{NewmanPenrose} the
Newman-Penrose formalism is applied to the non-symmetric fields of the
Einstein-Schr\"{o}dinger theory. In \S\ref{FieldEquations} the field
equations are expressed in tetrad form. In \S\ref{Upsilon} a
solution is derived for $\tGam^\alpha_{\sigma\beta}(\N_{\mu\nu})$
in the tetrad frame. In \S\ref{ClassicalEM} the theory is shown to
approximate ordinary general relativity and electromagnetism when a
cosmological constant from zero-point fluctuations is assumed.
In \S\ref{Monopole} the tetrad form of an exact electric monopole
solution is shown to approximate the Reissner-Nordstr\"{o}m solution
and to be of Petrov type-D.

\section{\label{NewmanPenrose}Application of the Newman-Penrose Formalism}
In the remainder of this paper we will assume $n\!=\!4$. Let us write
(\ref{gdef},\ref{fdef}) as
\begin{eqnarray}
\label{Mdef}
\Nbar^{\sigma\mu}&=&\frac{\rmN}{\rmg}\K^{\dashv\mu\sigma}
=g^{\sigma\mu}+f^{\sigma\mu}.
\end{eqnarray}
It is proven in \cite{Hlavaty} and in \ref{Proof} that if we do not have
$f^\sigma{_\mu}f^\mu{_\sigma}\!=det(f^{\mu}{_{\nu}})\!=\!0$,
then $\Nbar^{\sigma\mu}$ can be decomposed such that,
\begin{eqnarray}
~\Nbar^{\sigma\mu}
&=&\Nbar^{{a}{}{b}}e_{{a}}{^\sigma}e_{{b}}{^\mu},\\
\label{M}
~\Nbar^{{a}{}{b}} &=&
\pmatrix{
0&\!\!(1\!+\!\tanht)&0&0\cr
(1\!-\!\tanht)&0&0&0\cr
0&0&0&\!\!-(1\!+\!i\tant)\cr
0&0&\!\!-(1\!-\!i\tant)&0
},\\
\label{gtetrad}
~~g_{{a}{}{b}}
&=&~g^{{a}{}{b}}
=\Nbar^{({a}{}{b})}=
\pmatrix{
~0&1&0&0\cr
~1&0&0&0\cr
~0&0&0&\!\!\!-1\cr
~0&0&\!\!-1&0
},\\
\label{ftetrad}
-f_{{a}{}{b}}
&=&f^{{a}{}{b}}
=\Nbar^{[{a}{}{b}]}=
\pmatrix{
0&\!\tanht&0&0\cr
-\tanht&0&0&0\cr
0&0&0&\!\!\!\!-i\tant\cr
0&0&\!i\tant&0
}.
\end{eqnarray}
Here tetrad indices are indicated with Latin letters.
As with the usual Newman-Penrose formalism\cite{Newman},
the tetrads $e_{{a}}{^\sigma}$ and inverse tetrads $e^{{a}}{_\mu}$
both consist of two real vectors and two complex conjugate vectors,
and are related via raising and lowering of indices with
$g^{{a}{}{b}}$ and $g_{\sigma\mu}$,
\begin{eqnarray}
~l^\sigma\!&=&e_{{1}}{^\sigma}\,,\,
n^\sigma\!=e_{{2}}{^\sigma}\,,\,
m^\sigma\!=e_{{3}}{^\sigma}~~,\,
m^{\!*\,\sigma}\!=e_{{4}}{^\sigma},\\
~l_\sigma\!&=&e^{{2}}{_\sigma}\,,\,
n_\sigma\!=e^{{1}}{_\sigma}\,,\,
m_\sigma\!=-e^{{4}}{_\sigma}\,,\,
m^*_\sigma\!=-e^{{3}}{_\sigma},\\
\label{inverse}
\delta^\sigma_\mu
\!&=&e_{{a}}{^\sigma}e^{{a}}{_\mu}~~~~~~~~~,~
\delta^{{a}}_{{b}}=e_{{b}}{^\sigma}e^{{a}}{_\sigma},\\
\label{etilde}
~\dete\!&=&det(e^{{a}}{_\nu})
=\epsilon^{\alpha\beta\sigma\mu}l_\alpha n_\beta m_\sigma m^*_\mu\\
\dete^*&=&-\dete.
\end{eqnarray}
One difference from the usual Newman-Penrose formalism is that gauge freedom
is restricted so that only type III tetrad transformations can be used.
If $\Nbar^{\sigma\mu}$ is real, the scalars
``$\tant$'' (u grave) and ``$\tanht$'' (u check)
are real and are given by\cite{Hlavaty}
\begin{eqnarray}
\label{tantdef}
~~~\tant&=&\sqrt{\sqrt{\HD}-\ff/4}\,,\\
~~~\tanht&=&\sqrt{\sqrt{\HD}+\ff/4}\,,\\
~~~\HD&=&(\ff/4)^2-f/g,\\
~~~f&=&det(f_{\mu\nu})~~,~~g=det(g_{\mu\nu}),\\
f/g&=&-\tanht^2\tant^2,\\
\label{ff}
~~\ff&=&f^\sigma{_\mu}f^\mu{_\sigma}
=2(\tanht^2\!-\!\tant^2).
\end{eqnarray}
If $\Nbar^{\sigma\mu}$ is instead Hermitian,
things are unchanged except that
``$\tant$'' and ``$\tanht$'' are imaginary instead of real.

From (\ref{Mdef},\ref{M}),
the fundamental tensor of the Einstein-Schr\"{o}dinger theory is,
\begin{eqnarray}
\label{Ninverse}
\K^{\dashv{a}{}{b}} &=&
\frac{\rmgt}{\rmNt}\!\!
\pmatrix{
0&\!\!\!\!(1\!-\!\tanht)&0&0\cr
(1\!+\!\tanht)&0&0&0\cr
0&0&0&\!\!\!\!\!\!\!-(1\!-\!i\tant)\cr
0&0&\!\!\!\!\!\!\!-(1\!+\!i\tant)&0
}\!,\\
\label{N}
\N_{{b}{}{c}}&=&
\pmatrix{
0&\!\!\!\!\!\!(1\!-\!\tanht)\cosht/\cost&0&0\cr
(1\!+\!\tanht)\cosht/\cost&0&0&0\cr
0&0&0&\!\!\!\!\!\!-(1\!-\!i\tant)\cost/\cosht\cr
0&0&\!\!\!\!\!\!-(1\!+\!i\tant)\cost/\cosht&0
}\!,
\end{eqnarray}
where
\begin{eqnarray}
\label{cost}
~~~~~~~\cost&=&\frac{1}{\sqrt{1\!+\!\tant^2}}
=\sqrt{1\!-\!\sint^2}\,,\\
~~~~~~~\tant&=&\sint/\cost,\\
\label{cosht}
~~~~~~~\cosht&=&\frac{1}{\sqrt{1\!-\!\tanht^2}}
=\sqrt{1\!+\!\sinht^2}\,,\\
~~~~~~~\tanht&=&\sinht/\cosht,\\
\label{rmN0}
\rmNt&=&\sqrt{-det(\N_{ab})}
=\frac{i}{\cosht\cost},\\
\label{rmg0}
~\rmgt&=&\sqrt{-det(g_{ab})}=i,\\
\label{rmN}
~\rmN&=&\rmNt\,\dete
=\frac{i\dete}{\cost\cosht},\\
\label{rmg}
~~\rmg&=&\rmgt\,\dete=i\dete.
\end{eqnarray}
Note the correspondence of $\sint, \cost, \tant$ and $\sinht, \cosht, \tanht$
to circular and hyperbolic trigonometry functions.

Covariant derivative is done in the usual fashion,
\begin{eqnarray}
\label{covariantderivative}
\fl T^{{a}}{}{_{{b}|{c}}}
&=&e^{{a}}{}{_\sigma}e_{{b}}{^\mu}T^\sigma{_{\mu;\tau}}e_{{c}}{^\tau}
=T^{{a}}{}{_{{b},{c}}}
+\gamma^{{a}}{}{_{{d}{}{c}}}T^{{d}}{}{_{{b}}}
-\gamma^{{d}}{}{_{{b}{}{c}}}T^{{a}}{}{_{{d}}}.
\end{eqnarray}
For the spin coefficients we will follow the conventions
of Chandrasekhar\cite{Chandrasekhar},
\begin{eqnarray}
\label{gammadef}
\fl ~~~~~~\gamma_{{a}{}{b}{}{c}}
=\frac{1}{2}(\lambda_{{a}{}{b}{}{c}}
+\lambda_{{c}{}{a}{}{b}}-\lambda_{{b}{}{c}{}{a}})
=e_{{a}}{^\mu}e_{{b}\mu;\sigma}e_{{c}}{^\sigma},\\
\label{gammasymmetry}
\fl ~~~~~~\gamma_{{a}{}{b}{}{c}}
=-\gamma_{{b}{}{a}{}{c}}~~,~~
\gamma^{{a}}{_{{a}{}{c}}}=0,\\
\label{lambdadef}
\fl ~~~~~~\lambda_{{a}{}{b}{}{c}}
=(e_{{b}\sigma,\mu}
-e_{{b}\mu,\sigma})e_{{a}}{^\sigma}e_{{c}}{^\mu}
=e_{{b}\sigma,\mu}(e_{{a}}{^\sigma}e_{{c}}{^\mu}-e_{{a}}{^\mu}e_{{c}}{^\sigma})
=\gamma_{{a}{}{b}{}{c}}-\gamma_{{c}{}{b}{}{a}},\\
\label{lambdasymmetry}
\fl ~~~~~~\lambda_{{a}{}{b}{}{c}}
=-\lambda_{{c}{}{b}{}{a}},\\
\fl ~~~~~~\rho=\gamma_{{3}{}{1}{}{4}}
~~,~~\mu=\gamma_{{2}{}{4}{}{3}}
~~,~~\tau=\gamma_{{3}{}{1}{}{2}}
~~,~~\pi=\gamma_{{2}{}{4}{}{1}},\\
\fl ~~~~~~\kappa=\gamma_{{3}{}{1}{}{1}}~~,~~\sigma=\gamma_{{3}{}{1}{}{3}}
~~,~~\lambda=\gamma_{{2}{}{4}{}{4}}~~,~~\nu=\gamma_{{2}{}{4}{}{2}},\\
\fl ~~~~~~\epsilon~=(\gamma_{{2}{}{1}{}{1}}+\gamma_{{3}{}{4}{}{1}})/2
~~~~,~~\gamma=(\gamma_{{2}{}{1}{}{2}}+\gamma_{{3}{}{4}{}{2}})/2,\\
\fl ~~~~~~\alpha=(\gamma_{{2}{}{1}{}{4}}+\gamma_{{3}{}{4}{}{4}})/2
~~~~,~~\beta=(\gamma_{{2}{}{1}{}{3}}+\gamma_{{3}{}{4}{}{3}})/2.
%\fl \epsilon^*+\epsilon&=&\gamma_{{2}{}{1}{}{1}}~~,~~
%\epsilon^*-\epsilon=\gamma_{{4}{}{3}{}{1}},\\
%\fl \gamma^*+\gamma&=&\gamma_{{2}{}{1}{}{2}}~~,~~
%\gamma^*-\gamma=\gamma_{{4}{}{3}{}{2}},\\
%\fl \alpha^*+\beta&=&\gamma_{{2}{}{1}{}{3}}~~,~~
%\alpha^*-\beta=\gamma_{{4}{}{3}{}{3}}\\
%\fl ~~~~~~\Psi_0&=&-C_{{1}{}{3}{}{1}{}{3}},\\
%\fl ~~~~~~\Psi_1&=&-C_{{1}{}{2}{}{1}{}{3}},\\
%\fl ~~~~~~\Psi_2&=&-C_{{1}{}{3}{}{4}{}{2}},\\
%\fl ~~~~~~\Psi_3&=&-C_{{1}{}{2}{}{4}{}{2}},\\
%\fl ~~~~~~\Psi_4&=&-C_{{2}{}{4}{}{2}{}{4}}.
\end{eqnarray}
With these coefficients and with other tetrad quantities, complex conjugation
causes the exchange ${3}\nobreak\!\nobreak\rightarrow\nobreak\!\nobreak{4},
{4}\!\nobreak\rightarrow\nobreak\!{3}$.
As usual we may also define directional derivative operators,
\begin{eqnarray}
\fl~~~~~D&=&e_1{^\alpha}\frac{\partial}{\partial x^\alpha}~~~,~~~
\Delta=e_2{^\alpha}\frac{\partial}{\partial x^\alpha}~~~,~~~
\delta=e_3{^\alpha}\frac{\partial}{\partial x^\alpha}~~~,~~~
\delta^*=e_4{^\alpha}\frac{\partial}{\partial x^\alpha}.
\end{eqnarray}
\section{\label{FieldEquations}The Field Equations in Newman-Penrose Form}
Substituting (\ref{Ricciaddition}) into (\ref{para}) gives the field equations,
\begin{eqnarray}
\label{NPpara}
\fl\frac{8\pi G}{c^4}\!\left(T_{{b}{d}}
-\frac{1}{2}g_{{b}{d}}T^{{a}}_{{a}}\right)
&=&R_{{b}{}{d}}
+2\Lambda_b\Aphi_{[{b}|{d}]}
+\Lambda_b \N_{{b}{}{d}}+\Lambda_e g_{{b}{}{d}}\nonumber\\
\fl&&+\Upsilon^{{a}}_{\!{b}{}{d}|{a}}
-\Upsilon^{{a}}_{\!{a}({b}|{d})}
-\Upsilon^{{c}}_{\!{b}{}{a}}\Upsilon^{{a}}_{\!{c}{}{d}}
+\Upsilon^{{c}}_{\!{b}{}{d}}\Upsilon^{{a}}_{\!{c}{}{a}}.
\end{eqnarray}
Taking the symmetric and antisymmetric parts of this and rearranging gives,
\begin{eqnarray}
\label{NPparasymmetric}
\fl~~~~R_{{b}{}{d}}
&=&\frac{8\pi G}{c^4}\!\left(T_{{b}{d}}
-\frac{1}{2}g_{{b}{d}}T^{{a}}_{{a}}\right)
-\Lambda_b \N_{({b}{}{d})}-\Lambda_e g_{{b}{}{d}}\nonumber\\
\fl&&-\Upsilon^{{a}}_{\!({b}{}{d})|{a}}
+\Upsilon^{{a}}_{{a}{}({b}|{d})}
+\Upsilon^{{c}}_{\!({b}{}{a})}
\Upsilon^{{a}}_{\!({c}{}{d})}
+\Upsilon^{{c}}_{\![{b}{}{a}]}
\Upsilon^{{a}}_{\![{c}{}{d}]}
-\Upsilon^{{c}}_{\!({b}{}{d})}
\Upsilon^{{a}}_{{c}{}{a}},\\
\label{NPparaantisymmetric}
\fl \Lambda_b \N_{[{b}{}{d}]}&=&
2\Lambda_b\Aphi_{[{d}|{b}]}
-\Upsilon^{{a}}_{\![{b}{}{d}]|{a}}
+\Upsilon^{{c}}_{\!({b}{}{a})}
\Upsilon^{{a}}_{\![{c}{}{d}]}
+\Upsilon^{{c}}_{\![{b}{}{a}]}
\Upsilon^{{a}}_{\!({c}{}{d})}
-\Upsilon^{{c}}_{\![{b}{}{d}]}
\Upsilon^{{a}}_{{c}{}{a}}.
\end{eqnarray}
The usual Ricci identities will be valid if we define $\Phi_{ab}$ values in
terms of the right-hand side of (\ref{NPparasymmetric}),
\begin{eqnarray}
\fl~~~~\Phi_{00}=-R_{11}/2,~\Phi_{22}=-R_{22}/2,~\Phi_{02}
=-R_{33}/2,~\Phi_{20}=-R_{44}/2,\\
\fl~~~~\Phi_{01}=-R_{13}/2,~\Phi_{10}=-R_{14}/2,~\Phi_{12}
=-R_{23}/2,~\Phi_{21}=-R_{24}/2,\\
\fl~~~~\Phi_{11}=-(R_{12}+R_{34})/4,~\hat\Lambda=R/24=(R_{12}-R_{34})/12.
\end{eqnarray}

From (\ref{ftetrad},\ref{covariantderivative}), Ampere's law
(\ref{Ampere}) becomes,
\begin{eqnarray}
\fl\frac{4\pi}{c}\J^{{c}}
&=&f^{{b}{}{c}}{_{,{b}}}
+\gamma^{{b}}_{{a}{}{b}}f^{{a}{}{c}}
+\!\gamma^{{c}}_{{a}{}{b}}f^{{b}{}{a}},\\
\fl\frac{4\pi}{c}\J^{{2}}
&=&f^{{1}{}{2}}{_{,{1}}}
+\gamma^{{3}}_{{1}{}{3}}f^{{1}{}{2}}
+\gamma^{{4}}_{{1}{}{4}}f^{{1}{}{2}}
+\gamma^{{2}}_{{3}{}{4}}f^{{4}{}{3}}
+\gamma^{{2}}_{{4}{}{3}}f^{{3}{}{4}}\\
\fl&=&\NPD\tanht
-\rho^*\tanht-\rho\tanht-\rho i\tant+\rho^*i\tant\\
\label{Ampere1}
\fl&=&\NPD\tanht-\rho w-\rho^*w^*,\\
\fl\frac{4\pi}{c}\J^{{1}}
&=&f^{{2}{}{1}}{_{,{2}}}
+\gamma^{{3}}_{{2}{}{3}}f^{{2}{}{1}}
+\gamma^{{4}}_{{2}{}{4}}f^{{2}{}{1}}
+\gamma^{{1}}_{{3}{}{4}}f^{{4}{}{3}}
+\gamma^{{1}}_{{4}{}{3}}f^{{3}{}{4}}\\
\fl&=&- \NPDEL\tanht
-\mu\tanht-\mu^*\tanht+\mu^*i\tant-\mu i\tant\\
\label{Ampere2}
\fl&=&-\NPDEL\tanht-\mu w-\mu^*w^*,\\
\fl\frac{4\pi}{c}\J^{{4}}
&=&f^{{3}{}{4}}{_{,{3}}}
+\gamma^{{1}}_{{3}{}{1}}f^{{3}{}{4}}
+\gamma^{{2}}_{{3}{}{2}}f^{{3}{}{4}}
+\gamma^{{4}}_{{1}{}{2}}f^{{2}{}{1}}
+\gamma^{{4}}_{{2}{}{1}}f^{{1}{}{2}}\\
\fl&=&-i\NPdel\tant
-\pi^*i\tant+\tau i\tant+\tau\tanht+\pi^*\tanht\\
\label{Ampere3}
\fl&=&-i\NPdel\tant+\tau w+\pi^*w^*,
\end{eqnarray}
where
\begin{eqnarray}
\fl~~~w&=&\tanht+i\tant
\end{eqnarray}
%\pagebreak

The connection equations are easier to work with in contravariant
form (\ref{contravariant}) than in covariant form (\ref{JSconnection}).
Multiplying (\ref{contravariant}) by $\rmN/\!\rmg$ and using
(\ref{Ninverse},\ref{covariantderivative},\ref{sum},\ref{rmN0},\ref{rmg0})
gives
\begin{eqnarray}
\fl~~0=~\FO^{{c}{}{d}}_{{b}}
&=&\frac{\rmNt}{\rmgt}\left(\K^{\dashv{c}{}{d}}{}{_{,{b}}}
+\gamma^{{d}}_{{a}{}{b}}\K^{\dashv{c}{}{a}}
+\gamma^{{c}}_{{a}{}{b}}\K^{\dashv{a}{}{d}}
+\Upsilon^{{d}}_{\!{a}{}{b}}\K^{\dashv{c}{}{a}}
+\Upsilon^{{c}}_{\!{b}{}{a}}\K^{\dashv{a}{}{d}}
\right)\nonumber\\
\fl~~&&+\frac{8\pi}{3c}\left(
\J^{[{d}}\delta^{{c}]}_{{b}}
-\frac{1}{2}\J^{{a}}\N_{[{a}{b}]}\K^{\dashv{c}{d}}\right),\\
\label{NPcontravariant0}
\fl~~0=~\FO^{{1}{}{1}}_{{b}}
&=&~~\Upsilon^{{1}}_{\!{2}{}{b}}(1\!-\!\tanht)
+\Upsilon^{{1}}_{\!{b}{}{2}}(1\!+\!\tanht),\\
\fl~~0=~\FO^{{2}{}{2}}_{{b}}
&=&~~\Upsilon^{{2}}_{\!{1}{}{b}}(1\!+\!\tanht)
+\Upsilon^{{2}}_{\!{b}{}{1}}(1\!-\!\tanht),\\
\fl~~0=~\FO^{{3}{}{3}}_{{b}}
&=&-\Upsilon^{{3}}_{\!{4}{}{b}}(1\!-\!i\tant)
-\Upsilon^{{3}}_{\!{b}{}{4}}(1\!+\!i\tant),\\
\fl~~0=\!{^\pm\!}\FO^{{1}{}{2}}_{{b}}
&=&\mp\tanht_{,{b}}
+\OMPU\left(-\frac{(\rmNt\,){_{,{b}}}}{\rmNt}
+{^\pm\!}\Upsilon^{{2}}_{\!{2}{}{b}}
+{^\pm\!}\Upsilon^{{1}}_{\!{b}{}{1}}\right)\nonumber\\
\fl~~&&+\frac{8\pi}{3c}\left(
\pm \J^{[{2}}\delta^{{1}]}_{{b}}
-\frac{1}{2}\J^{{a}}\N_{[{a}{b}]}\cost\cosht\OMPU\right)\\
\fl&=&\OMPU\left(\mp\tanht_{,{b}}\cosht^2
-\tant\tant_{,{b}}\cost^2
+{^\pm\!}\Upsilon^{{2}}_{\!{2}{}{b}}
+{^\pm\!}\Upsilon^{{1}}_{\!{b}{}{1}}\right)\nonumber\\
\fl~~&&+\frac{8\pi}{3c}\left(
\frac{\pm 1}{\OPMU}\J^{[{2}}\delta^{{1}]}_{{b}}
+i\tant\frac{\OMPU}{(1\!+\!\tant^2)}\J^{[{4}}\delta^{{3}]}_{{b}}
\!\right)\!,\\
\fl~~0=\!{^\pm\!}\FO^{{3}{}{4}}_{{b}}
&=&\pm i\tant_{,{b}}
+\OMPIV\left(\frac{(\rmNt\,){_{,{b}}}}{\rmNt}
-{^\pm\!}\Upsilon^{{4}}_{\!{4}{}{b}}
-{^\pm\!}\Upsilon^{{3}}_{\!{b}{}{3}}\right)\nonumber\\
\fl~~&&+\frac{8\pi}{3c}\left(
\pm \J^{[{4}}\delta^{{3}]}_{{b}}
+\frac{1}{2}\J^{{a}}\N_{[{a}{b}]}\cost\cosht\OMPIV\right)\\
\fl&=&\OMPIV\left(\pm i\tant_{,{b}}\cost^2
-\tanht\tanht_{,{b}}\cosht^2
-{^\pm\!}\Upsilon^{{4}}_{\!{4}{}{b}}
-{^\pm\!}\Upsilon^{{3}}_{\!{b}{}{3}}\right)\nonumber\\
\fl~~&&+\frac{8\pi}{3c}\left(
\frac{\pm 1}{\OPMIV}\J^{[{4}}\delta^{{3}]}_{{b}}
+\tanht\frac{\OMPIV}{(1\!-\!\tanht^2)}\J^{[{2}}\delta^{{1}]}_{{b}}
\!\right)\!,\\
\fl~~0=\!{^\pm\!}\FO^{{2}{}{4}}_{{b}}
&=&\gamma_{{3}{}{1}{}{b}}(-\OPMU+\OMPIV)
+{^\pm\!}\Upsilon^{{4}}_{\!{1}{}{b}}\OPMU
-{^\pm\!}\Upsilon^{{2}}_{\!{b}{}{3}}\OMPIV
\pm\frac{8\pi}{3c}
\J^{[{4}}\delta^{{2}]}_{{b}}\\
\fl&=&\mp\gamma_{{3}{}{1}{}{b}}w
+{^\pm\!}\Upsilon^{{4}}_{\!{1}{}{b}}\OPMU
-{^\pm\!}\Upsilon^{{2}}_{\!{b}{}{3}}\OMPIV
\pm\frac{8\pi}{3c}\J^{[{4}}\delta^{{2}]}_{{b}},\\
\fl~~0=\!{^\pm\!}\FO^{{1}{}{3}}_{{b}}
&=&\gamma_{{2}{}{4}{}{b}}(\OMPU-\OPMIV)
+{^\pm\!}\Upsilon^{{3}}_{\!{2}{}{b}}\OMPU
-{^\pm\!}\Upsilon^{{1}}_{\!{b}{}{4}}\OPMIV
\pm\frac{8\pi}{3c}
\J^{[{3}}\delta^{{1}]}_{{b}}\\
\label{NPcontravariant1}
\fl&=&\mp\gamma_{{2}{}{4}{}{b}}w
+{^\pm\!}\Upsilon^{{3}}_{{2}{}{b}}\OMPU
-{^\pm\!}\Upsilon^{{1}}_{{b}{}{4}}\OPMIV
\pm\frac{8\pi}{3c}\J^{[{3}}\delta^{{1}]}_{{b}},
\end{eqnarray}
To save space in the equations above we are using the notation,
\begin{eqnarray}
\label{plusminus}
\fl ~~~~~~~~{^-}\FO^{{d}{}{c}}_{{b}}
&=&~{^+}\FO^{{c}{}{d}}_{{b}}
=~\FO^{{c}{}{d}}_{{b}}~~,~~
{^-}\Upsilon^{{d}}_{{c}{}{b}}
={^+}\Upsilon^{{d}}_{{b}{}{c}}
=\Upsilon^{{d}}_{{b}{}{c}}.
\end{eqnarray}
\section{\label{Upsilon}An Exact Solution for the Connection Addition in the
Tetrad Frame}
The connection equations (\ref{NPcontravariant0}-\ref{NPcontravariant1})
can be solved by forming linear combinations of them where all of the
$\Upsilon^{{a}}_{\!{b}{}{c}}$ terms cancel except for
the desired one. The required linear combinations are listed in
\ref{LinearCombinations} and the calculations are done in
\ref{Calculation}. The result, split into symmetric and antisymmetric
components is below,
\begin{eqnarray}
\fl \Upsilon^{{2}}_{\!({1}{}{2})}&=&~~~\cosht^2\tanht\NPD\tanht
-\frac{4\pi\cosht^2\tanht}{3c}\J^{{2}},\\
\fl \Upsilon^{{1}}_{\!({1}{}{2})}&=&~~~\cosht^2\tanht\NPDEL\tanht
+\frac{4\pi\cosht^2\tanht}{3c}\J^{{1}},\\
\fl \Upsilon^{{4}}_{\!({3}{}{4})}&=&-\cost^2\tant\NPdel\tant
+\frac{4\pi\cost^2i\tant}{3c}\J^{{4}},\\
\fl\Upsilon^{{1}}_{\!({1}{}{1})}
&=&\tant\NPD\tant\cost^2-\tanht\NPD\tanht\cosht^2
+\frac{4\pi\tanht\cosht^2}{3c}\J^{{2}},\\
\fl\Upsilon^{{2}}_{\!({2}{}{2})}
&=&\tant\NPDEL\tant\cost^2-\tanht\NPDEL\tanht\cosht^2
-\frac{4\pi\tanht\cosht^2}{3c}\J^{{1}},\\
\fl\Upsilon^{{3}}_{\!({3}{}{3})}
&=&\tant\NPdel\tant\cost^2-\tanht\NPdel\tanht\cosht^2
-\frac{4\pi i\tant\cost^2}{3c}\J^{{4}},\\
\fl \Upsilon^{{2}}_{\!({1}{}{1})}
&=&\Upsilon^{{1}}_{\!({2}{}{2})}
=\Upsilon^{{3}}_{\!({4}{}{4})}=0,\\
\fl \Upsilon^{{2}}_{\!({2}{}{3})}
&=&~~~\frac{i\tant}{2}(\NPdel\tanht\cosht^2-i\NPdel\tant\cost^2)
-\frac{2\pi i\tant\cost^2}{3c}\J^{{4}},\\
\fl \Upsilon^{{1}}_{\!({1}{}{3})}
&=&-\!\frac{i\tant}{2}(\NPdel\tanht\cosht^2+i\NPdel\tant\cost^2)
-\frac{2\pi i\tant\cost^2}{3c}\J^{{4}},\\
\fl \Upsilon^{{3}}_{\!({1}{}{3})}
&=&-\frac{\tanht}{2}(\NPD\tanht\cosht^2+i\NPD\tant\cost^2)
+\frac{2\pi \tanht\cosht^2}{3c}\J^{{2}},\\
\fl \Upsilon^{{3}}_{\!({2}{}{3})}
&=&-\frac{\tanht}{2}(\NPDEL\tanht\cosht^2-i\NPDEL\tant\cost^2)
-\frac{2\pi \tanht\cosht^2}{3c}\J^{{1}},\\
\fl \Upsilon^{{4}}_{\!({1}{}{2})}
&=&-\frac{~\tanht\cosht^2}{2}\left(
\NPdel\tanht\frac{\cosht^2}{\cost^2}
+\tau w-\pi^* w^*\right)\!,\\
\fl \Upsilon^{{2}}_{\!({3}{}{4})}
&=&-\frac{i\tant\cost^2}{2}\left(
i\NPD\tant\frac{\cost^2}{\cosht^2}
+\rho w-\rho^* w^*\right)\!,\\
\fl \Upsilon^{{1}}_{\!({4}{}{3})}
&=&-\frac{i\tant\cost^2}{2}\left(i\NPDEL\tant\frac{\cost^2}{\cosht^2}
-\mu w+\mu^* w^*\right)\!,\\
\fl \Upsilon^{{2}}_{\!({1}{}{3})}
&=&~\frac{\kappa w\tanht}{\Zht}~~,~
\Upsilon^{{1}}_{\!({2}{}{4})}
=-\frac{\nu w\tanht}{\Zht},\\
\fl \Upsilon^{{4}}_{\!({1}{}{3})}
&=&\frac{\sigma wi\tant}{\Zt}~~,~
\Upsilon^{{3}}_{\!({2}{}{4})}
=-\frac{\lambda wi\tant}{\Zt},\\
\fl \Upsilon^{{4}}_{\!({1}{}{1})}
&=&~\frac{\kappa w^2}{\Zht}~~,~
\Upsilon^{{3}}_{\!({2}{}{2})}
=-\frac{\nu w^2}{\Zht},\\
\fl \Upsilon^{{2}}_{\!({3}{}{3})}
&=&\frac{\sigma w^2}{\Zt}~~~,~
\Upsilon^{{1}}_{\!({4}{}{4})}
=-\frac{\lambda w^2}{\Zt},\\
\fl \Upsilon^{{2}}_{\![{1}{}{2}]}&=&-\cosht^2\NPD\tanht
+\frac{4\pi\cosht^2}{3c}\J^{{2}},\\
\fl \Upsilon^{{1}}_{\![{1}{}{2}]}&=&-\cosht^2\NPDEL\tanht
-\frac{4\pi\cosht^2}{3c}\J^{{1}},\\
\fl \Upsilon^{{4}}_{\![{3}{}{4}]}&=&- i\cost^2\NPdel\tant
-\frac{4\pi\cost^2}{3c}\J^{{4}},\\
\fl \Upsilon^{{2}}_{\![{2}{}{3}]}
&=&~~~\frac{1}{2}(\NPdel\tanht\cosht^2-i\NPdel\tant\cost^2)
-\frac{2\pi\cost^2}{3c}\J^{{4}},\\
\fl \Upsilon^{{1}}_{\![{1}{}{3}]}
&=&-\!\frac{1}{2}(\NPdel\tanht\cosht^2+i\NPdel\tant\cost^2)
-\frac{2\pi\cost^2}{3c}\J^{{4}},\\
\fl \Upsilon^{{3}}_{\![{1}{}{3}]}
&=&~~~\frac{1}{2}(\NPD\tanht\cosht^2+i\NPD\tant\cost^2)
-\frac{2\pi\cosht^2}{3c}\J^{{2}},\\
\fl \Upsilon^{{3}}_{\![{2}{}{3}]}
&=&-\!\frac{1}{2}(\NPDEL\tanht\cosht^2-i\NPDEL\tant\cost^2)
-\frac{2\pi\cosht^2}{3c}\J^{{1}},\\
\fl \Upsilon^{{4}}_{\![{1}{}{2}]}
&=&~~~\frac{\cosht^2}{2}\left(
\NPdel\tanht\frac{\cosht^2}{\cost^2}
+\tau w-\pi^* w^*\right)\!,\\
\fl \Upsilon^{{2}}_{\![{3}{}{4}]}
&=&~~~\frac{\cost^2}{2}\left(
i\NPD\tant\frac{\cost^2}{\cosht^2}
+\rho w-\rho^* w^*\right)\!,\\
\fl \Upsilon^{{1}}_{\![{4}{}{3}]}
&=&-\!\frac{\cost^2}{2}\left(
i\NPDEL\tant\frac{\cost^2}{\cosht^2}
-\mu w+\mu^* w^*\right)\!,\\
\fl \Upsilon^{{2}}_{\![{1}{}{3}]}
&=&-\!\frac{\kappa w}{\Zht}~~,~~
\Upsilon^{{1}}_{\![{2}{}{4}]}
=-\frac{\nu w}{\Zht},\\
\fl \Upsilon^{{4}}_{\![{1}{}{3}]}
&=&~~~\frac{\sigma w}{\Zt}~~,~~
\Upsilon^{{3}}_{\![{2}{}{4}]}
=~~\frac{\lambda w}{\Zt},
\end{eqnarray}
where
\begin{eqnarray}
\label{Zdef}
\fl ~~~~~~~~~~
\Zt&=&[\OPMIV^2\OPMU+\OMPIV^2\OMPU]/2=1+2i\tanht\tant-\tant^2,\\
\label{Zdef2}
\fl ~~~~~~~~~~
\Zht&=&[\OPMU^2\OPMIV+\OMPU^2\OMPIV]/2=1+2i\tanht\tant+\tanht^2.
\end{eqnarray}
As an error check, it is easy to verify that these results agree with
(\ref{selftransplant}) and (\ref{JScontractionsymmetric}),
\begin{eqnarray}
\fl \Upsilon^{{a}}_{\!({b}{}{a})}
\!&=&\!\tant\tant_{,{b}}\cost^2\!-\!\tanht\tanht_{,{b}}\cosht^2
\!+\!\frac{8\pi}{3c}\left(\tanht\cosht^2\delta^{[{1}}_{{b}}\J^{{2}]}
-i\tant\cost^2\delta^{[{3}}_{{b}}\J^{{4}]}\right)\\
\fl &=&-\frac{(\rmgt\,)_{,{b}}}{\rmgt}
+\frac{(\rmNt\,)_{,{b}}}{\rmNt}
\!+\!\frac{4\pi}{3c}\frac{\rmgt}{\rmNt}
\J^{{a}}\N_{[{a}{b}]},\\
\fl\Upsilon^{{a}}_{\![{b}{}{a}]}\!&=&0.
\end{eqnarray}
%\pagebreak
\section{\label{ClassicalEM}Approximation of Classical General Relativity
and Electromagnetism}
In \cite{Shifflett,Shifflett3} it is shown that this theory closely approximates
ordinary general relativity and electromagnetism.
Here we will confirm this by deriving
some of the results in \cite{Shifflett,Shifflett3} with tetrad methods.

We will make much use of the small-skew approximation\cite{Shifflett},
\begin{eqnarray}
\label{smallskew}
~~~~~~~~~|f^\alpha{_\beta}|\ll 1.
\end{eqnarray}
It is a widely accepted technique, and is used heavily in research
on this topic. We will refer to equations as being accurate to
order $f^1$ or $f^2$ etc., meaning that higher order terms such as
$f^\alpha{_\beta}f^\beta{_\alpha}f^\sigma{_\mu}$ are being ignored.
To a limited extent we will also use the approximation of small rates
of change and small spatial curvatures,
\begin{eqnarray}
\label{smallderivative}
|f^\alpha{_{\beta;\nu}}/f^\alpha{_\beta}|&\ll& \sqrt{\Lambda_b},\\
\label{smallcurvature}
~~~~~~|C_{\sigma\mu\alpha\rho}|&\ll& \Lambda_b,
\end{eqnarray}
where $C_{\sigma\mu\alpha\rho}$ is the Weyl tensor. The symbols $||$ mean
the largest measurable component for some standard spherical or cartesian
coordinate system. If an equation has a tensor term which can be neglected in
one coordinate system, it can be neglected in any coordinate system, so it is
only necessary to prove it in one coordinate system. We will see that the
approximations above are satisfied to such an extreme degree that it is simply
not necessary to define them more rigorously.

Let us consider worst-case values of
(\ref{smallskew},\ref{smallderivative},\ref{smallcurvature}).
We will assume that $\Lambda_e$ is caused by zero-point fluctuations with a
cutoff wave-number\cite{Sakharov,Padmanabhan,Padmanabhan2,Ashtekar,Smolin}
\begin{eqnarray}
\label{cutoff}
k_c\!=\!C_c/l_P~~~~,~~~~C_c\sim 1,
\end{eqnarray}
where $l_P=\sqrt{\hbar G/c^3}=1.6\times 10^{-33}$cm is the Planck length. Then
assuming all of the known fundamental particles we have\cite{Sahni,Shifflett},
\begin{eqnarray}
\label{Lambdab}
\fl~~~~~~~~~\Lambda_b\approx-\Lambda_e\sim C_z k_c^4 l_P^2\sim{C_c^4C_z}/{l_P^2}
\sim 10^{66}\,{\rm cm}^{-2}~~~~,~~~~ C_z\sim 60/2\pi?
\end{eqnarray}
For a charged particle with $f^1{_0}=Q/r^2$,
applying the conversion to cgs units (\ref{cgs}) shows that
$|f^1{_0}|=1$ would occur at the radius
\begin{eqnarray}
\label{redef}
\fl~~~~~~~~~r_e&=&\sqrt{|Q|}
=\sqrt{\left|\,e\sqrt{\!\frac{\,\lower2pt\hbox{$-2G$}}{c^4\Lambda_b}}\right|}
=\sqrt{l_P \sqrt{\frac{\lower2pt\hbox{$2e^2$}}{c\hbar\Lambda_b}}}
=\frac{l_P}{C_c}\left(\frac{2\alpha}{C_z}\right)^{1/4}\!\!\sim 10^{-33}{\rm cm}
\end{eqnarray}
where $\alpha =e^2/\hbar c\approx 1/137$ is the fine structure constant.
For atomic radii near the Bohr radius
($a_0=\hbar^2/m_e e^2=5.3\times\! 10^{-9}{\rm cm}$) we have,
\begin{eqnarray}
\label{Bohrskew}
~~~~~~~~~~~~~~|f^1{_0}|&\sim& r_e^2/a_0^2\sim 10^{-50},\\
\label{Bohrderiv}
|f^1{_{0;1}}/\sqrt{\Lambda_b}\,f^1{_0}|
&\sim& 2/\sqrt{\Lambda_b}\,a_0\sim 10^{-24}.
\end{eqnarray}
For the smallest radii probed by high energy particle physics
experiments ($10^{-17}{\rm cm}$),
\begin{eqnarray}
\label{highenergyskew}
~~~~~~~~~~~~~~|f^1{_0}|&\sim& r_e^2/(10^{-17})^2\sim 10^{-32},\\
\label{highenergyderiv}
|f^1{_{0;1}}/\sqrt{\Lambda_b}\,f^1{_0}|
&\sim& 2/\sqrt{\Lambda_b}\,10^{-17}\sim 10^{-16}.
\end{eqnarray}
The fields at this radius are larger than near any macroscopic charged object,
and are also larger than the strongest plane-wave fields. We must also consider
rates of change for the highest energy gamma rays ($10^{20}$eV) where
%$k\!=\!E/\hbar c\!=5\times 10^{24}\rm{rad/cm}$ so that,
%Assuming the field is localized to a volume $1/k^3$ and
%using (\ref{cgs2},\ref{Lambdab}) gives
\begin{eqnarray}
%\fl|f^1{_0}|\sim \sqrt{4\pi k^3 E}\sqrt{\!\frac{\,-2G}{c^4\Lambda_b}}
%\sim\sqrt{4\pi k^4\hbar c }\sqrt{\!\frac{\,-2G}{c^4\Lambda_b}}
%&\sim&\frac{4\pi k^2 l_P^2}{\sqrt{60}}\sim 10^{-16}\\
\label{gammaderiv}
|f^1{_{0;1}}/\sqrt{\Lambda_b}\,f^1{_0}|
&\sim& E/\hbar c\sqrt{\Lambda_b}\sim 10^{-8}.
\end{eqnarray}
The largest observable values of the Weyl tensor might be expected to occur
near the Schwarzschild radius, $r_s\!\nobreak=\nobreak\!2Gm/c^2$, of black
holes, where it takes on values around $r_s/r^3$. However, since
the lightest black holes have the smallest Schwarzschild radius,
they will create the largest value of $r_s/r_s^3=1/r_s^2$. The lightest black
hole that we can expect to observe would be of about one solar mass, where
\begin{eqnarray}
\label{blackholecurvature}
\frac{C_{0101}}{\Lambda_b}
&\sim&\frac{1}{\Lambda_br^2_s}
=\frac{1}{\Lambda_b}\left(\frac{c^2}{2Gm_\odot}\right)^2\!\sim\!10^{-77}.
\end{eqnarray}
Clearly the approximations
(\ref{smallskew},\ref{smallderivative},\ref{smallcurvature})
are extremely accurate. This is particularly true for the small-skew
approximation because terms with higher powers of $f{^\sigma}_\mu$ are usually
two powers higher than leading order terms, so that from (\ref{highenergyskew})
they will be $<\!10^{-64}$ of the leading order terms.
\bigbreak

The tetrad formalism allows the small-skew approximation to be
stated somewhat more rigorously as $|\tant|\ll 1,~|\tanht|\ll 1$.
From (\ref{tantdef}-\ref{ff}), a charged particle will have
$\tanht\approx \Q/r^2,\tant=0$.
From (\ref{cost},\ref{cosht},\ref{ff},\ref{N},\ref{gtetrad},\ref{ftetrad})
we have, to second order in $\tant~{\rm and}~\tanht$,
\begin{eqnarray}
\label{coshtocost}
\fl~~~~~~~~\cosht/\cost
&\approx&~~~\,1+\tanht^2/2+\tant^2/2
=~~1+\tanht^2-\ff/4,\\
\label{costocosht}
\fl~~~~-\cost/\cosht
&\approx&-1+\tanht^2/2+\tant^2/2
=-1+\tant^2+\ff/4,\\
\fl~~~~~~\N_{({a}{}{b})}
&=&\pmatrix{
0&\cosht/\cost&0&0\cr
\cosht/\cost&0&0&0\cr
0&0&0&\!\!\!-\cost/\cosht\cr
0&0&\!\!\!-\cost/\cosht&0
}
\label{Nbarapprox}
\approx g_{{a}{}{b}}+f_{{a}}{^{{c}}}f_{{c}{}{b}}
-\frac{1}{4}g_{{a}{}{b}}\ff,\\
\fl~~~~~~\N_{[{a}{}{b}]}
&=&\pmatrix{
0&-\!\tanht\cosht/\cost&0&0\cr
\tanht\cosht/\cost&0&0&0\cr
0&0&0&i\tant\cost/\cosht\cr
0&0&-i\tant\cost/\cosht&0
}
\label{Nhatapprox}
\approx f_{{a}{}{b}}.
\end{eqnarray}
These $n=4$ results match the order $f^2$ approximations
derived in \cite{Shifflett},
\begin{eqnarray}
\label{approximateNbar}
\N_{(\sigma\mu)}&\approx& g_{\sigma\mu}+{f_\sigma}^\nu f_{\nu\mu}
-\frac{1}{2(n\!-\!2)}g_{\sigma\mu}\ff,\\
\label{approximateNhat}
\N_{[\sigma\mu]}&\approx& f_{\sigma\mu}.
\end{eqnarray}
The next higher order terms of (\ref{coshtocost},\ref{costocosht}) will be two
orders higher in $\tant~{\rm and}~\tanht$ than the leading order terms.
This confirms that the next higher order terms in
(\ref{approximateNbar},\ref{approximateNhat}) will be two orders higher in
$f^\mu{_\nu}$ than the leading order terms, and from (\ref{highenergyskew})
these terms must be $<\!10^{-64}$ of the leading order terms.
Also, while we are mostly ignoring the special case
$f^\sigma{_\mu}f^\mu{_\sigma}\!=det(f^{\mu}{_{\nu}})\!=\!0$
in this paper, it is easy to show from \ref{Proof} that for this case,
(\ref{approximateNbar},\ref{approximateNhat}) are exact instead of approximate.

In \ref{Check} it is shown that to second order in $\tant~{\rm and}~\tanht$,
the exact $n\!=\!4$ solution for $\Upsilon^{{a}}_{\!{b}{}{c}}$
in \S\ref{Upsilon} matches the order $f^2$
approximation derived in \cite{Antoci3,Shifflett},
\begin{eqnarray}
\fl~~~~~~{\Upsilon}^\alpha_{\!(\sigma\mu)}
\label{upsilonsymmetric}
&\approx&f^\tau{_{(\sigma}}f_{\mu)}{^\alpha}{_{;\tau}}
+f^{\alpha\tau}f_{\tau(\sigma;\mu)}
+\frac{1}{4(n\!-\!2)}(\ff_,{^\alpha}g_{\sigma\mu}
-2\ff_{,(\sigma}\delta^\alpha_{\mu)})\nonumber\\
\fl&&+\frac{4\pi}{c(n\!-\!2)}\J^\rho\left(f^\alpha{_\rho}\,g_{\sigma\mu}
+\frac{2}{(n\!-\!1)}f_{\rho(\sigma}\delta^\alpha_{\mu)}\right),\\
\label{upsilonantisymmetric}
\fl~~~~~~{\Upsilon}^\alpha_{\![\sigma\mu]}
&\approx&\frac{1}{2}(f_{\sigma\mu;}{^\alpha}
+f^\alpha{_{\mu;\sigma}}
-f^\alpha{_{\sigma;\mu}})
+\frac{8\pi}{c(n\!-\!1)}\J_{[\sigma}\delta^\alpha_{\mu]},\\
\label{upsiloncontracted}
\fl~~~~~~\Upsilon^\alpha_{\alpha\sigma}
&\approx&\frac{-1}{2(n\!-\!2)}\ff_{,\sigma}
+\frac{8\pi}{c(n\!-\!1)(n\!-\!2)}\J^\alpha f_{\alpha\sigma}.
\end{eqnarray}
The tetrad version of (\ref{upsilonsymmetric}-\ref{upsilonantisymmetric})
in \ref{Check} differs from the
exact solution in \S\ref{Upsilon} only by the factors
$\cost$,$\cosht$,$\Zt$,$\Zht$,
and from (\ref{coshtocost},\ref{costocosht},\ref{Zdef},\ref{Zdef2})
these factors will induce terms which are two
orders higher in $\tant$ and $\tanht$ than the leading order terms.
This confirms that the next higher order terms in
(\ref{upsilonsymmetric}-\ref{upsiloncontracted}) will be two orders higher
in $f^\mu{_\nu}$ than the leading order terms, and
from (\ref{highenergyskew}) these terms must be $<\!10^{-64}$
of the leading order terms.

Substituting (\ref{approximateNbar}) into (\ref{Einstein}) and using
(\ref{ff}) gives the order $f^2$ Einstein equations,
\begin{eqnarray}
\label{Einstein2}
\fl~~~~~~~~~\tG_{\sigma\mu}
&\approx&\frac{8\pi G}{c^4}T_{\sigma\mu}
-\Lambda_b\left({f_\sigma}^\nu f_{\nu\mu}
\!-\!\frac{1}{4}g_{\sigma\mu}f^{\rho\nu}f_{\nu\rho}\right)
+\Lambda\left(\frac{n}{2}-1\right)g_{\sigma\mu}.
\end{eqnarray}
With the conversion to cgs units (\ref{cgs}), the second term
in (\ref{Einstein2}) is the ordinary electromagnetic energy-momentum tensor.
By substituting
(\ref{upsilonsymmetric},\ref{upsilonantisymmetric},\ref{upsiloncontracted})
into (\ref{Ricciaddition},\ref{genEinstein}), one can
derive\cite{Shifflett,Shifflett3} an order $f^2$ approximation of
$\tG_{\sigma\mu}$ in terms of the ordinary Einstein tensor
$G_{\sigma\mu}\nobreak
\!=\!\nobreak R_{\sigma\mu}\nobreak\!-\!\nobreak (1/2)g_{\sigma\mu}R$.
However, without actually doing the calculation, it is easy to see that
$\tG_{\sigma\mu}$ and $G_{\sigma\mu}$ can differ only by second
order terms such as $f^\tau{_{(\sigma}}f_{\mu)}{^\alpha}{_{;\tau;\alpha}}$
and $f^\nu{_{\sigma;\alpha}}f^\alpha{_{\mu;\nu}}$. This result applies with or
without charge currents since $4\pi \J^\rho/c\!=\!f^{\tau\rho}{_{;\tau}}$
from (\ref{Ampere}).
Therefore from (\ref{gammaderiv},\ref{highenergyderiv}), such additional terms
must be $<\!10^{-16}$ of the ordinary electromagnetic term.
%Therefore (\ref{Einstein2}) is clearly a close approximation to the Einstein
%equations of ordinary general relativity and electromagnetism.

Combining
(\ref{paraantisymmetric},\ref{approximateNhat},\ref{Ricciaddition},
\ref{upsilonantisymmetric}) gives, to order $f^2$,
\begin{eqnarray}
\fl 2\Lambda_b\Aphi_{[\sigma,\mu]}\!+\!\Lambda_bf_{\sigma\mu}
&\approx&-(\tR_{[\sigma\mu]}
+\tGam^\alpha_{\alpha[\sigma,\mu]})
\approx-\Upsilon^\alpha_{\![\sigma\mu];\alpha}\\
\fl~~~~&\approx&-\frac{1}{2}(f_{\sigma\mu;}{^\alpha}
\!+\!f^\alpha{_{\mu;\sigma}}\!-\!f^\alpha{_{\sigma;\mu}}){_{;\alpha}}
-\frac{8\pi}{c(n\!-\!1)}\J_{[\sigma,\mu]}\\
\fl &\approx&
-\!\frac{3}{2}f_{[\sigma\mu,\alpha];}{^\alpha}
+2f^\alpha{_{[\sigma;\mu];\alpha}}
-\frac{8}{c(n\!-\!1)}\J_{[\sigma,\mu]},\\
\fl &\approx&
-\!\frac{3}{2}f_{[\sigma\mu,\alpha];}{^\alpha}
+4f^\alpha{_{[\sigma;[\mu];\alpha]}}
+\frac{8\pi}{c} \J_{[\sigma,\mu]}
-\frac{8\pi}{c(n\!-\!1)}\J_{[\sigma,\mu]},\\
\label{antisymmetricpreliminary}
\fl ~~~~~~f_{\sigma\mu}\approx2\Aphi_{[\mu,\sigma]}
&+&\dual_{[\tau,\alpha]}\varepsilon_{\sigma\mu}{^{\tau\alpha}}
+\frac{4}{\Lambda_b}f^\alpha{_{[\sigma;[\mu];\alpha]}}
+\frac{8\pi(n\!-\!2)}{\Lambda_b c(n\!-\!1)}\,\J_{[\sigma,\mu]},
\end{eqnarray}
where
\begin{eqnarray}
\label{vartheta}
\fl~~~~~~~~~~~~~f_{[\sigma\mu,\alpha]}=\frac{-2\Lambda_b}{3}
\,\dual_\tau\varepsilon^\tau{_{\sigma\mu\alpha}}~~,~~
\dual_\tau=\frac{1}{4\Lambda_b}
f_{[\sigma\mu,\alpha]}\varepsilon_\tau{^{\sigma\mu\alpha}}.
\end{eqnarray}
From (\ref{highenergyderiv},\ref{Ampere}), the last term of
(\ref{antisymmetricpreliminary}) can only contribute
$<\!10^{-32}$ of $f_{\sigma\mu}$.
From (\ref{blackholecurvature},\ref{Einstein2},\ref{highenergyskew})
the $4f^\alpha{_{[\sigma;[\mu];\alpha]}}/\Lambda_b$ term can only contribute
$<\!10^{-77}$ of $f_{\sigma\mu}$ because of the antisymmetrized second
derivative. Ignoring these terms and taking the divergence of
(\ref{antisymmetricpreliminary}) using (\ref{Ampere}),
the $\dual_{[\tau,\alpha]}\varepsilon_{\sigma\mu}{^{\tau\alpha}}$ term
drops out and we get a close approximation to Maxwell's equations,
\begin{eqnarray}
\label{Maxwell}
\fl~~~~~~~~~~~~~F_{\sigma\mu;}{^\sigma}&=&2\Aphi_{[\mu,\sigma];}{^\sigma}
\approx\frac{\lower2pt\hbox{$4\pi$}}{c}\J_\mu ,\\
\label{Faraday}
\fl~~~~~~~~~~~~F_{[\sigma\mu,\nu]}&=&2\Aphi_{[\mu,\sigma,\nu]}=0
~~~~~~~~{\rm (from~the~definition~F_{\sigma\mu}=2\Aphi_{[\mu,\sigma]}).}
\end{eqnarray}

The $\dual_{[\tau,\alpha]}\varepsilon_{\sigma\mu}{^{\tau\alpha}}$ term of
(\ref{antisymmetricpreliminary}) can also be neglected from
(\ref{highenergyderiv},\ref{vartheta}). However, it is interesting to consider
the case where (\ref{smallderivative}) is not satisfied, but where there are no
extreme spatial curvatures (\ref{smallcurvature}) so that
$4f^\alpha{_{[\sigma;[\mu];\alpha]}}/\Lambda_b$ is still negligible.
Then, taking the curl of (\ref{antisymmetricpreliminary}),
the $2\Aphi_{[\mu,\sigma]}$ and $j_{[\sigma,\mu]}$ terms drop
out and we see that the additional vector field $\dual_\rho$ obeys
a form of the Proca equation,
\begin{eqnarray}
\Lambda_b\dual_\rho
\label{rawProca}
\approx-\dual_{[\rho,\nu];}{^{\nu}}.
\end{eqnarray}
This equation suggests the possibility of a $\dual_\rho$ particle, and this is
discussed in detail in \S6-\S7 of \cite{Shifflett}. There it is shown that if a
$\dual_\rho$ particle does result from (\ref{rawProca}), it would apparently
have a negative energy. However, the other odd feature of this particle is that
from (\ref{rawProca},\ref{cutoff},\ref{Lambdab}), its
%A minimum frequency $\dual_\rho$ wave is a ``stationary'' wave of the form
%$\dual_\sigma\!\nobreak=\nobreak\!(0,\overrightarrow{\epsilon} sin(\omega t))$
%where $\overrightarrow{\epsilon}$ is a constant 3-vector, and
%from (\ref{cutoff},\ref{Lambdab})
minimum frequency $\omega\!=\!\sqrt{2\Lambda_b}\,c\!=\!\sqrt{2C_z}\,cC_c^2/l_P$
would exceed the zero-point cutoff frequency $ck_c\!=\!cC_c/l_P$, and we assume
this would prevent the particle from existing. Whether the cutoff of zero-point
fluctuations is caused by a discreteness of
spacetime near the Planck length\cite{Padmanabhan,Padmanabhan2,Ashtekar,Smolin}
or by some other effect, we simply assume that this cutoff would also apply to
real fundamental particles\cite{Sakharov}. Comparing the two frequency
expressions, we see that this argument only applies if
\begin{eqnarray}
\label{limit}
C_c>1/\sqrt{2C_z}\,,
\end{eqnarray}
where $C_c$ and $C_z$ are defined by (\ref{cutoff},\ref{Lambdab}). Since the
prediction of a negative energy particle would probably be inconsistent with
reality, this theory should be approached cautiously when considering it with
values of $C_c$ and $C_z$ which do not satisfy (\ref{limit}).

%\pagebreak
\section{\label{Monopole}An Electric Monopole Solution}
Here we assume $T_{\sigma\mu}=0,~\J^\rho=0$, which is the
Einstein-Schr\"{o}dinger equivalent of electro-vac general relativity.
The Newman-Penrose tetrads
of the electric monopole solution derived in \cite{Shifflett} are similar to
those of the Reissner-Nordstr\"{o}m solution\cite{Chandrasekhar},
except for the $\cosht$ factors,
\begin{eqnarray}
\label{ltetrad}
\fl~~ e_{{1}\alpha}=l_\alpha&=&(1,-1/a\cosht,0,0)~~~~~~~~~~~~~~,~~
e_{{1}}{^\alpha}=l^\alpha=(1/a\cosht,1,0,0),\\
\label{ntetrad}
\fl~~ e_{{2}\alpha}=n_\alpha&=&\frac{1}{2}(a\cosht,1,0,0)~~~~~~~~~~~~~~~~~,~~
e_{{2}}{^\alpha}=n^\alpha=\frac{1}{2}(1,-a\cosht,0,0),\\
\label{mtetrad}
\fl~~ e_{{3}\alpha}=m_\alpha&=&-r\sqrt{\cosht/2}\,
(0,0,1,i\,{\rm sin}\,\theta)~~,~~
e_{{3}}{^\alpha}=m^\alpha=\frac{\lower2pt\hbox{$1$}}{r\sqrt{2\cosht}}
(0,0,1,i\,{\rm csc}\,\theta),
\end{eqnarray}
where from \cite{Shifflett} and (\ref{Lambdab},\ref{cgs}),
\begin{eqnarray}
\fl ~~\tant&=&0~~,~~\sint=0~~,~~\cost=1,\\
\fl ~~\tanht&=&\frac{\sinht}{\cosht}
=\frac{\Q}{\cosht\,r^2},\\
\label{shat}
\fl ~~\sinht&=&\frac{\Q}{r^2},\\
\label{chat}
\fl ~~\cosht&=&\frac{1}{\sqrt{1-\tanht^2}}
=\sqrt{1+\sinht^2}=\sqrt{1+\frac{\Q^2}{r^4}}\,,\\
\label{afunction}
\fl ~~a&=&1-\frac{2m}{r}-\frac{\Lambda_br^2}{3}
-\frac{\Lambda_eV}{r}
=1-\frac{2m}{r}-\frac{\Lambda r^2}{3}
-(\Lambda_b\!-\!\Lambda)
\!\left(\frac{\Q^2}{2r^2}-\!\frac{\Q^4}{40r^6}~...\right),\\
\fl ~~V&=&\int{\sqrt{r^4+\Q^2}\,dr}
=\frac{1}{3}
\!\left[r\sqrt{r^4\!+\Q^2}
-\Q^{3/2}F\!\left(\!2\,{\rm arctan}
\!\left(\!\!\frac{\sqrt{\Q^{\mathstrut\!\!}}}{r}
\right)\!,\!\frac{\pi}{4}\right)\!\right]\\
\fl ~~&=&\int r^2
\left(1+\frac{\Q^2}{2r^4}-\frac{\Q^4}{8r^8}~...\right)dr
=\frac{r^3}{3}
-\frac{\Q^2}{2r}+\frac{\Q^4}{40r^5}~...~,\\
\label{Lambdavalue}
\fl ~\Lambda_b&\approx&-\Lambda_e\sim \pm 10^{66}{\rm cm}^{-2}
~~,~~\Lambda\sim 10^{-56}{\rm cm}^{-2}
~~,~~\Lambda/\Lambda_b\sim 10^{-122},\\
\label{Qvalue}
\fl ~~\Q&=&e\sqrt{\frac{-2G}{c^4\Lambda_b}}
\sim\sqrt{\mp 1}\,\times 10^{-66}{\rm cm}^2.
\end{eqnarray}
In (\ref{afunction}), the term $-\Lambda_b\Q^2/2r^2\!=\!e^2G/c^4r^2$
matches a term appearing in the Reissner-Nordstr\"{o}m solution, and
the remaining $\Q$ terms are negligible for ordinary radii.

The nonzero tetrad derivatives are,
\begin{eqnarray}
\fl e_{{1}1,1}&=&-\left(\frac{1}{a\cosht}\right)'~~~,~~~
e_{{2}0,1}=\frac{(a\cosht)'}{2}~~~,~~~
e_{{3}3,2}=-r\sqrt{\cosht/2}\,i\,{\rm cos}\,\theta\,,\\
\fl e_{{3}2,1}&=&-\sqrt{\cosht/2}-\frac{r\cosht'}{2\sqrt{2\cosht}}
=\frac{-\cosht^2+\sinht^2}{\sqrt{2\cosht}\,\cosht}
=\frac{-1}{\sqrt{2\cosht}\,\cosht}~~~,~~~
e_{{3}3,1}=e_{{3}2,1}\,i\,{\rm sin}\,\theta\,.
\end{eqnarray}
From these and (\ref{lambdadef}), the $\lambda_{{a}{}{b}{}{c}}$
coefficients are
\begin{eqnarray}
\fl\lambda_{{a}{}{1}{}{b}}
&=&e_{{1}1,1}(e_{{a}}{^1}e_{{b}}{^1}
                   \!-e_{{a}}{^1}e_{{b}}{^1})\!=0,\\
\fl\lambda_{{2}{}{2}{}{1}}
&=&e_{{2}0,1}(e_{{2}}{^0}e_{{1}}{^1}
                   \!-e_{{2}}{^1}e_{{1}}{^0})
%=\frac{(a\cosht)'}{2},\\
=\frac{(a\cosht)'}{2},\\
\fl\lambda_{{1}{}{2}{}{3}}
&=&e_{{2}0,1}(e_{{1}}{^0}e_{{3}}{^1}
                   \!-e_{{1}}{^1}e_{{3}}{^0})\!=0,\\
\fl\lambda_{{2}{}{2}{}{3}}
&=&e_{{2}0,1}(e_{{2}}{^0}e_{{3}}{^1}
                   \!-e_{{2}}{^1}e_{{3}}{^0})\!=0,\\
\fl\lambda_{{3}{}{2}{}{4}}
&=&e_{{2}0,1}(e_{{3}}{^0}e_{{4}}{^1}
                   \!-e_{{3}}{^1}e_{{4}}{^0})\!=0,\\
\fl\lambda_{{1}{}{3}{}{2}}
&=&e_{{3}0,1}(e_{{1}}{^0}e_{{2}}{^1}
                   \!-e_{{1}}{^1}e_{{2}}{^0})\!=0,\\
\fl\lambda_{{2}{}{3}{}{3}}
&=&-e_{{3}2,1}e_{{2}}{^1}e_{{3}}{^2}
         \!-e_{{3}3,1}e_{{2}}{^1}e_{{3}}{^3}\!=0,\\
\fl\lambda_{{2}{}{4}{}{3}}
&=&-e_{{4}2,1}e_{{2}}{^1}e_{{3}}{^2}
         \!-e_{{4}3,1}e_{{2}}{^1}e_{{3}}{^3}
=-2\left(\frac{-1}{\sqrt{2\cosht}\,\cosht}\right)
 \!\!\left(\frac{-a\cosht}{2}\right)\frac{1}{r\sqrt{2\cosht}}
=-\frac{a}{2r\cosht},\\
\fl\lambda_{{4}{}{4}{}{1}}
&=&e_{{4}2,1}e_{{4}}{^2}e_{{1}}{^1}
        \!+e_{{4}3,1}e_{{4}}{^3}e_{{1}}{^1}\!=0,\\
\fl\lambda_{{4}{}{3}{}{1}}
&=&e_{{3}2,1}e_{{4}}{^2}e_{{1}}{^1}
        \!+e_{{3}3,1}e_{{4}}{^3}e_{{1}}{^1}
=2\left(\frac{-1}{\sqrt{2\cosht}\,\cosht}\right)
\frac{1}{r\sqrt{2\cosht}}
=-\frac{1}{r\cosht^2},\\
\fl\lambda_{{3}{}{3}{}{4}}
&=&e_{{3}3,2}(e_{{3}}{^3}e_{{4}}{^2}
                   \!-e_{{3}}{^2}e_{{4}}{^3})
=2(-r\sqrt{\cosht/2}\,i\,{\rm cos}\,\theta)
\left(\frac{i\,{\rm csc}\,\theta}{r\sqrt{2\cosht}}\right)
\frac{1}{r\sqrt{2\cosht}}
=\frac{{\rm cot}\,\theta}{r\sqrt{2\cosht}}\,.
\end{eqnarray}
From (\ref{gammadef}), the spin coefficients are similar to those of the
Reissner-Nordstr\"{o}m solution\cite{Chandrasekhar}, except for
the $\cosht$ factors,
\begin{eqnarray}
\fl~~~\rho
&=&\gamma_{{3}{}{1}{}{4}}=\lambda_{{4}{}{3}{}{1}}
=-\frac{1}{r\cosht^2},\\
\fl~~~\mu
&=&\gamma_{{2}{}{4}{}{3}}=\lambda_{{2}{}{4}{}{3}}
=-\frac{a}{2r\cosht},\\
\fl~~~\beta
&=&\frac{1}{2}(\gamma_{{2}{}{1}{}{3}}\!+\!\gamma_{{3}{}{4}{}{3}})
=\frac{1}{2}\lambda_{{3}{}{3}{}{4}}
\!=\!\frac{{\rm cot}\,\theta}{2r\sqrt{2\cosht}},\\
\fl~~~\alpha
&=&\frac{1}{2}(\gamma_{{2}{}{1}{}{4}}\!+\!\gamma_{{3}{}{4}{}{4}})
=\frac{1}{2}\lambda_{{3}{}{4}{}{4}}
\!=\!\frac{-{\rm cot}\,\theta}{2r\sqrt{2\cosht}},\\
\fl~~~\gamma
&=&\frac{1}{2}(\gamma_{{2}{}{1}{}{2}}\!+\!\gamma_{{3}{}{4}{}{2}})
=\frac{1}{2}\lambda_{{2}{}{2}{}{1}}\!=\!\frac{(a\cosht)'}{4},\\
\fl~~~\epsilon
&=&\frac{1}{2}(\gamma_{{2}{}{1}{}{1}}\!+\!\gamma_{{3}{}{4}{}{1}})\!=\!0,\\
\fl~~~\tau
&=&\gamma_{{3}{}{1}{}{2}}=0,\\
\fl~~~\pi
&=&\gamma_{{2}{}{4}{}{1}}=0,\\
\fl~~~\kappa
&=&\gamma_{{3}{}{1}{}{1}}=0,\\
\fl~~~\sigma
&=&\gamma_{{3}{}{1}{}{3}}=0,\\
\fl~~~\lambda
&=&\gamma_{{2}{}{4}{}{4}}=0,\\
\fl~~~\nu
&=&\gamma_{{2}{}{4}{}{2}}=0.
%\fl~~~\tau
%&=&\gamma_{{3}{}{1}{}{2}}=0,\\
%\fl~~~\pi
%&=&\gamma_{{2}{}{4}{}{1}}=0.
%\fl~~~\kappa
%&=&\gamma_{{3}{}{1}{}{1}}=0~~,~~
%\sigma
%=\gamma_{{3}{}{1}{}{3}}=0~~,~~
%\lambda
%=\gamma_{{2}{}{4}{}{4}}=0~~,~~
%\nu
%=\gamma_{{2}{}{4}{}{2}}=0.
\end{eqnarray}

The type-D classification of this solution is evident because
$\kappa\!=\!\sigma\!=\!\lambda\!=\!\nu\!=\!\epsilon\!=\!0$
and from the Weyl tensor components calculated with MAPLE,
\begin{eqnarray}
\fl~~~\Psi_2&=&
-\frac{1}{\cosht}\left(1+\frac{2\Q^2}{r^4}\right)
\!\left(\frac{m}{r^3}+\frac{\Lambda_e V}{2r^3}-\frac{\Lambda_e\cosht}{6}\right)
+\frac{\Lambda_e \Q^2}{6r^4}+\frac{\Q^2}{2\cosht r^6}\,,\\
\fl~~~\Psi_0&=&\Psi_1=\Psi_3=\Psi_4=0.
\end{eqnarray}
The electromagnetic vector potential from \cite{Shifflett} and
(\ref{ltetrad}-\ref{mtetrad}) completes the solution,
\begin{eqnarray}
\label{phi}
\fl~~~~~~~\Aphi_{{a}}&=&(\Aphi_0/a\cosht,\Aphi_0/2,0,0),\\
\fl~~~~~~~\Aphi_0&=&\frac{\Q}{r}
\!\left(1-\!\frac{4\Lambda}{3\Lambda_b}\right)
+\frac{\Q m}{\Lambda_b r^4}
+\!\left(1\!-\!\frac{\Lambda}{\Lambda_b}\right)
\!\!\left(\frac{2\Q^3}{5r^5}-\frac{\Q^5}{30r^9}~...\right).
\end{eqnarray}
Here all terms except $\Q/r$ are negligible for ordinary radii,
assuming (\ref{Lambdavalue},\ref{Qvalue}).

\pagebreak
From
(\ref{Mdef},\ref{M},\ref{ltetrad}-\ref{mtetrad},\ref{shat}-\ref{afunction}),
the tetrad solution matches the solution in \cite{Shifflett},
\begin{eqnarray}
\label{Mprod1}
\fl \Nbar^{\sigma\mu}
&=&e^{\sigma}{_{{a}}}\Nbar^{{a}{}{b}}e_{{b}}{^\mu}\\
\fl &=&e^{\sigma}{_{{a}}}\!
\pmatrix{
0&\!\!1\!+\!\tanht&0&0\cr
1\!-\!\tanht&0&0&0\cr
0&0&0&\!\!-1\cr
0&0&\!\!\!\!\!-1&0
}\!\!\!
\pmatrix{
\frac{1}{a\cosht}&1&0&0\cr
\frac{1^{\vphantom{1}}}{2}&-\frac{a\cosht}{2}&0&0\cr
0&0&\frac{1}{r\sqrt{2\cosht}}&~~\frac{i\,{\rm csc}\,\theta}{r\sqrt{2\cosht}}\cr
0&0&\frac{1}{r\sqrt{2\cosht}}&-\frac{i\,{\rm csc}\,\theta}{r\sqrt{2\cosht}}
}\!\\
\fl &=&\!\pmatrix{
\frac{1}{a\cosht}&\!\!\frac{1}{2}&0&0\cr
1&\!\!-\frac{a\cosht}{2}&0&0\cr
0&0&\frac{1}{r\sqrt{2\cosht}}&\frac{1}{r\sqrt{2\cosht}}\cr
0&0&\frac{i\,{\rm csc}\,\theta}{r\sqrt{2\cosht}}
\,&-\frac{i\,{\rm csc}\,\theta}{r\sqrt{2\cosht}}\,
}\!\!\!
\pmatrix{
\!\frac{(1\!+\!\tanht)}{2}&-\frac{(1\!+\!\tanht)a\cosht}{2}&0&0\cr
\!\frac{(1\!-\!\tanht)}{a\cosht}&{(1\!-\!\tanht)}\atop{\vphantom{a}}&0&0\cr
0&0&-\frac{1}{r\sqrt{2\cosht}}&~~\frac{i\,{\rm csc}\,\theta}{r\sqrt{2\cosht}}\cr
0&0&-\frac{1}{r\sqrt{2\cosht}}&-\frac{i\,{\rm csc}\,\theta}{r\sqrt{2\cosht}}
}\!\\
\fl &=&\frac{1}{\cosht}\pmatrix{
1/a&-\sinht&0&0\cr
\sinht&-a\cosht^2&0&0\cr
0&0&\!\!-1/r^2&0\cr
0&0&0&\!\!\!-1/r^2{\rm sin}^2\theta
}\!,\\
\fl~g^{\sigma\mu}&=&\Nbar^{(\sigma\mu)}
=\frac{1}{\cosht}\pmatrix{
1/a&0&0&0\cr
0&-a\cosht^2&0&0\cr
0&0&\!\!-1/r^2&0\cr
0&0&0&\!\!\!-1/r^2{\rm sin}^2\theta
},\\
\fl~g_{\sigma\mu}&=&
~~~~~~~~~~~~~\cosht\pmatrix{
a&0&0&0\cr
0&-1/a\cosht^2&0&0\cr
0&0&\!\!-r^2&0\cr
0&0&0&\!\!\!-r^2{\rm sin}^2\theta
},\\
\fl~f^{\sigma\mu}&=&\Nbar^{[\sigma\mu]}
=\frac{1}{\cosht}\pmatrix{
\,0&-\sinht&0&0\cr
\sinht&0&0&0\cr
0&0&0&0\cr
0&0&0&0
}\!,\\
\fl~f_{\sigma\mu}&=&\Nbar_{[\sigma\mu]}
=\frac{1}{\cosht}\pmatrix{
0&\sinht&0&0\cr
-\sinht&0&0&0\cr
0&0&0&0\cr
0&0&0&0
},\\
\fl~\N_{\sigma\mu}&=&\frac{\rmN}{\rmg}\Nbar^{\dashv\,T}_{\sigma\mu}
=\pmatrix{
a\cosht^2&\sinht&0&0\cr
-\sinht&-1/a&0&0\cr
0&0&\!\!-r^2&0\cr
0&0&0&\!\!\!-r^2{\rm sin}^2\theta
}\!.
\end{eqnarray}
Also, from (\ref{Mprod1},\ref{rmN0}-\ref{rmg},\ref{etilde}) we have,
\begin{eqnarray}
\fl~~~~ 1/\dete&=&det(e_{{a}}{^\nu})=i\,{\rm csc}\,\theta/\cosht r^2,\\
\fl~~~~~~~\dete&=&det(e^{{a}}{_\nu})=-i\cosht r^2{\rm sin}\,\theta,\\
\fl~~ \rmN&=&\rmNt\,\dete
=i\dete/\cosht\cost=r^2{\rm sin}\,\theta,\\
\fl~~~\rmg&=&\rmgt\,\dete=i\dete
=\cosht r^2{\rm sin}\,\theta.
\end{eqnarray}

This solution reduces to the Papapetrou type I solution\cite{Papapetrou} for
$\Lambda_e\!=\!0,\Lambda_b\!=\!\Lambda$, and to the Schwarzschild solution
for $\Q=0$. Note from (\ref{Qvalue},\ref{redef},\ref{chat}) that for an
elementary charge, ``$\cosht\,$'' diverges from one near
$r_e\nobreak=\nobreak\sqrt{|\Q|}\sim 10^{-33}$cm, and it goes to zero there
if $\Q$ is imaginary. This $r_e$ surface may lie inside or outside of the
Schwarzschild radius depending on the charge/mass ratio.

\section{\label{Conclusions}Conclusions}

The Einstein-Schr\"{o}dinger theory is modified to include a large cosmological
constant caused by zero-point fluctuations. This extrinsic cosmological constant
which multiplies the symmetric metric is assumed to be nearly cancelled by
Schr\"{o}dinger's bare cosmological constant which multiplies the nonsymmetric
fundamental tensor, resulting in a total cosmological constant which is
consistent with measurement. Sections \S 1 and \S 5 demonstrate that this theory
closely approximates ordinary general relativity and electromagnetism,
confirming the results in \cite{Shifflett,Shifflett3} using different methods.
This is corroborated by the close approximation of the electric monopole solution
to the Reissner-Nordstr\"{o}m solution, and by its Petrov type-D classification.

The Einstein-Schr\"{o}dinger theory is very workable in Newman-Penrose
form. The presented solution of the connection equations is fairly simple and
is applicable for all cases except
$f^\sigma{_\mu}f^\mu{_\sigma}\!=\!det(f^{\mu}{_{\nu}})\!=\!0$.
It is very compatible with symbolic algebra programs such as REDUCE or MAPLE.
Given the amenability of Newman-Penrose methods to type-D solutions,
this paper might be useful in finding a charged rotating solution to
the Einstein-Schr\"{o}dinger theory, either in its original or modified form.

%\bigskip\\
\appendix

\section{\label{LinearCombinations}Linear Combinations to Solve for the
Connection Addition}
In the following linear combinations of equations
(\ref{NPcontravariant0}-\ref{NPcontravariant1}), the right-hand-side
$\Upsilon^{{a}}_{{b}{}{c}}$ terms cancel,
\begin{eqnarray}
\fl {^\pm\!}\Upsilon^{{2}}_{\!{1}{}{2}}
\!&=&\!{^\pm\!}\Upsilon^{{2}}_{\!{1}{}{2}}
+\frac{1}{2}\left(~~{^\pm\!}\FO^{{1}{}{2}}_{{1}}-\FO^{{2}{}{2}}_{{2}}
-{^\pm\!}\FO^{{2}{}{1}}_{{1}}\frac{\OMPU}{\OPMU}\right)\!,\\
\fl {^\pm\!}\Upsilon^{{1}}_{\!{1}{}{2}}
\!&=&\!{^\pm\!}\Upsilon^{{1}}_{\!{1}{}{2}}
+\frac{1}{2}\left(~~{^\pm\!}\FO^{{1}{}{2}}_{{2}}-\FO^{{1}{}{1}}_{{1}}
-{^\pm\!}\FO^{{2}{}{1}}_{{2}}\frac{\OMPU}{\OPMU}\right)\!,\\
\fl {^\pm\!}\Upsilon^{{4}}_{\!{3}{}{4}}
\!&=&\!{^\pm\!}\Upsilon^{{4}}_{\!{3}{}{4}}
+\frac{1}{2}\left(-{^\pm\!}\FO^{{3}{}{4}}_{{3}}+\FO^{{4}{}{4}}_{{4}}
+{^\pm\!}\FO^{{4}{}{3}}_{{3}}\frac{\OMPIV}{\OPMIV}\right)\!,\\
\fl ~\Upsilon^{{1}}_{\!{1}{}{1}}
\!&=&\!\Upsilon^{{1}}_{\!{1}{}{1}}
+\frac{1}{2}(-\FO^{{2}{}{1}}_{{1}}
-\FO^{{1}{}{2}}_{{1}}+\FO^{{2}{}{2}}_{{2}}),\\
\fl ~\Upsilon^{{2}}_{\!{2}{}{2}}
\!&=&\!\Upsilon^{{2}}_{\!{2}{}{2}}
+\frac{1}{2}(-\FO^{{1}{}{2}}_{{2}}
-\FO^{{2}{}{1}}_{{2}}+\FO^{{1}{}{1}}_{{1}}),\\
\fl ~\Upsilon^{{3}}_{\!{3}{}{3}}
\!&=&\!\Upsilon^{{3}}_{\!{3}{}{3}}
+\frac{1}{2}(~~\FO^{{4}{}{3}}_{{3}}
+\FO^{{3}{}{4}}_{{3}}-\FO^{{4}{}{4}}_{{4}}),\\
\fl ~\Upsilon^{{2}}_{\!{1}{}{1}}
\!&=&\!\Upsilon^{{2}}_{{1}{}{1}}-\frac{1}{2}\FO^{{2}{}{2}}_{{1}},\\
\fl ~\Upsilon^{{1}}_{\!{2}{}{2}}
\!&=&\!\Upsilon^{{1}}_{{2}{}{2}}-\frac{1}{2}\FO^{{1}{}{1}}_{{2}},\\
\fl ~\Upsilon^{{3}}_{\!{4}{}{4}}
\!&=&\!\Upsilon^{{3}}_{{4}{}{4}}+\frac{1}{2}\FO^{{3}{}{3}}_{{4}},\\
\fl {^\pm\!}\Upsilon^{{2}}_{\!{2}{}{3}}
\!&=&\!{^\pm\!}\Upsilon^{{2}}_{\!{2}{}{3}}
+\frac{1}{2}\left(~~{^\pm\!}\FO^{{2}{}{4}}_{{2}}-{^\pm\!}\FO^{{4}{}{1}}_{{1}}
-{^\pm\!}\FO^{{1}{}{2}}_{{3}}\frac{\OPMIV}{\OMPU}\right),\\
\fl {^\pm\!}\Upsilon^{{1}}_{\!{1}{}{3}}
\!&=&\!{^\pm\!}\Upsilon^{{1}}_{\!{1}{}{3}}
+\frac{1}{2}\left(-{^\pm\!}\FO^{{4}{}{2}}_{{2}}+{^\pm\!}\FO^{{1}{}{4}}_{{1}}
-{^\pm\!}\FO^{{2}{}{1}}_{{3}}\frac{\OPMIV}{\OPMU}\right),\\
\fl {^\pm\!}\Upsilon^{{3}}_{\!{1}{}{3}}
\!&=&\!{^\pm\!}\Upsilon^{{3}}_{\!{1}{}{3}}
+\frac{1}{2}\left(~~{^\pm\!}\FO^{{4}{}{2}}_{{4}}-{^\pm\!}\FO^{{2}{}{3}}_{{3}}
+{^\pm\!}\FO^{{3}{}{4}}_{{1}}\frac{\OMPU}{\OMPIV}\right),\\
\fl {^\pm\!}\Upsilon^{{3}}_{\!{2}{}{3}}
\!&=&\!{^\pm\!}\Upsilon^{{3}}_{\!{2}{}{3}}
+\frac{1}{2}\left(~~{^\pm\!}\FO^{{4}{}{1}}_{{4}}-{^\pm\!}\FO^{{1}{}{3}}_{{3}}
+{^\pm\!}\FO^{{3}{}{4}}_{{2}}\frac{\OPMU}{\OMPIV}\right),\\
\fl {^\pm\!}\Upsilon^{{4}}_{\!{1}{}{2}}
\!&=&\!{^\pm\!}\Upsilon^{{4}}_{\!{1}{}{2}}
+\frac{1}{2}\left(-{^\pm\!}\FO^{{2}{}{4}}_{{2}}\frac{\OPMIV}{\OPMU}
-{^\pm\!}\FO^{{4}{}{1}}_{{1}}\frac{\OMPIV}{\OPMU}
-{^\pm\!}\FO^{{1}{}{2}}_{{3}}\frac{(1\!+\!\tant^2)}{(1\!-\!\tanht^2)}
\right),\\
\fl {^\pm\!}\Upsilon^{{2}}_{\!{3}{}{4}}
\!&=&\!{^\pm\!}\Upsilon^{{2}}_{\!{3}{}{4}}
+\frac{1}{2}\left(~~{^\pm\!}\FO^{{4}{}{2}}_{{4}}\frac{\OPMU}{\OPMIV}
+{^\pm\!}\FO^{{2}{}{3}}_{{3}}\frac{\OMPU}{\OPMIV}
+{^\pm\!}\FO^{{3}{}{4}}_{{1}}\frac{(1\!-\!\tanht^2)}{(1\!+\!\tant^2)}
\right),\\
\fl {^\pm\!}\Upsilon^{{1}}_{\!{4}{}{3}}
\!&=&\!{^\pm\!}\Upsilon^{{1}}_{\!{4}{}{3}}
+\frac{1}{2}\left(~~{^\pm\!}\FO^{{3}{}{1}}_{{3}}\frac{\OMPU}{\OMPIV}
+{^\pm\!}\FO^{{1}{}{4}}_{{4}}\frac{\OPMU}{\OMPIV}
+{^\pm\!}\FO^{{4}{}{3}}_{{2}}\frac{(1\!-\!\tanht^2)}{(1\!+\!\tant^2)}
\right),\\
%\end{eqnarray}
%\begin{eqnarray}
\fl {^\pm\!}\Upsilon^{{2}}_{\!{1}{}{3}}
\!&=&\!{^\pm\!}\Upsilon^{{2}}_{\!{1}{}{3}}+\frac{1}{2\Zht}\left(
-{^\pm\!}\FO^{{4}{}{2}}_{{1}}(1\!-\!\tanht^2)
+{^\pm\!}\FO^{{2}{}{4}}_{{1}}\OMPU^2
-\FO^{{2}{}{2}}_{{3}}\OPMIV\OPMU\right),\\
\fl {^\pm\!}\Upsilon^{{1}}_{\!{2}{}{4}}
\!&=&\!{^\pm\!}\Upsilon^{{1}}_{\!{2}{}{4}}+\frac{1}{2\Zht}\left(
-{^\pm\!}\FO^{{3}{}{1}}_{{2}}(1\!-\!\tanht^2)
+{^\pm\!}\FO^{{1}{}{3}}_{{2}}\OPMU^2
-\FO^{{1}{}{1}}_{{4}}\OMPIV\OMPU\right),\\
\fl {^\pm\!}\Upsilon^{{4}}_{\!{1}{}{3}}
\!&=&\!{^\pm\!}\Upsilon^{{4}}_{\!{1}{}{3}}+\frac{1}{2\Zt}\left(
~{^\pm\!}\FO^{{4}{}{2}}_{{3}}(1\!+\!\tant^2)
-{^\pm\!}\FO^{{2}{}{4}}_{{3}}\OPMIV^2
+\FO^{{4}{}{4}}_{{1}}\OMPIV\OMPU\right),\\
\fl {^\pm\!}\Upsilon^{{3}}_{\!{2}{}{4}}
\!&=&\!{^\pm\!}\Upsilon^{{3}}_{\!{2}{}{4}}+\frac{1}{2\Zt}\left(
~{^\pm\!}\FO^{{3}{}{1}}_{{4}}(1\!+\!\tant^2)
-{^\pm\!}\FO^{{1}{}{3}}_{{4}}\OMPIV^2
+\FO^{{3}{}{3}}_{{2}}\OPMIV\OPMU\right),\\
\fl ~\Upsilon^{{4}}_{\!{1}{}{1}}
\!&=&\!\Upsilon^{{4}}_{\!{1}{}{1}}+\frac{1}{2\Zht}\left(
-\FO^{{2}{}{4}}_{{1}}\OPIV\OPU
-\FO^{{4}{}{2}}_{{1}}\OMIV\OMU
-\FO^{{2}{}{2}}_{{3}}(1\!+\!\tant^2)\right),\\
\fl ~\Upsilon^{{3}}_{\!{2}{}{2}}
\!&=&\!\Upsilon^{{3}}_{\!{2}{}{2}}+\frac{1}{2\Zht}\left(
-\FO^{{1}{}{3}}_{{2}}\OMIV\OMU
-\FO^{{3}{}{1}}_{{2}}\OPIV\OPU
-\FO^{{1}{}{1}}_{{4}}(1\!+\!\tant^2)\right),\\
\fl ~\Upsilon^{{2}}_{\!{3}{}{3}}
\!&=&\!\Upsilon^{{2}}_{\!{3}{}{3}}+\frac{1}{2\Zt}\left(
~~\FO^{{4}{}{2}}_{{3}}\OPIV\OPU
+\FO^{{2}{}{4}}_{{3}}\OMIV\OMU
+\FO^{{4}{}{4}}_{{1}}(1\!-\!\tanht^2)\right),\\
\fl ~\Upsilon^{{1}}_{\!{4}{}{4}}
\!&=&\!\Upsilon^{{1}}_{\!{4}{}{4}}+\frac{1}{2\Zt}\left(
~~\FO^{{3}{}{1}}_{{4}}\OMIV\OMU
+\FO^{{1}{}{3}}_{{4}}\OPIV\OPU
+\FO^{{3}{}{3}}_{{2}}(1\!-\!\tanht^2)\right).
\end{eqnarray}
\section{\label{Calculation}Calculation of the Exact Connection Addition}
Performing the linear combinations in \ref{LinearCombinations} and using
(\ref{Ampere1},\ref{Ampere2},\ref{Ampere3}) and the notation
(\ref{plusminus},\ref{Zdef}) gives
%\bigskip
\begin{eqnarray}
\fl {^\pm\!}\Upsilon^{{2}}_{{1}{}{2}}
&=&\mp \frac{\NPD\tanht}{\OPMU}
\pm\frac{4\pi}{3c\OPMU}\J^{{2}},\\
\fl {^\pm\!}\Upsilon^{{1}}_{{1}{}{2}}
&=&\mp \frac{\NPDEL\tanht}{\OPMU}
\mp\frac{4\pi}{3c\OPMU}\J^{{1}},\\
\fl {^\pm\!}\Upsilon^{{4}}_{{3}{}{4}}
&=&\mp \frac{i \NPdel\tant}{\OPMIV}
\mp\frac{4\pi}{3c\OPMIV}\J^{{4}},\\
\fl~\Upsilon^{{1}}_{{1}{}{1}}
&=&\frac{\NPD\rmNt}{\rmNt}
+\frac{4\pi\tanht\cosht^2}{3c}\J^{{2}}
=\tant\NPD\tant\cost^2-\tanht\NPD\tanht\cosht^2
+\frac{4\pi\tanht\cosht^2}{3c}\J^{{2}},\\
\fl~\Upsilon^{{2}}_{{2}{}{2}}
&=&\frac{\NPDEL\rmNt}{\rmNt}
-\frac{4\pi\tanht\cosht^2}{3c}\J^{{1}}
=\tant\NPDEL\tant\cost^2-\tanht\NPDEL\tanht\cosht^2
-\frac{4\pi\tanht\cosht^2}{3c}\J^{{1}},\\
\fl~\Upsilon^{{3}}_{{3}{}{3}}
&=&\frac{\NPdel\rmNt}{\rmNt}
-\frac{4\pi i\tant\cost^2}{3c}\J^{{4}}
=\tant\NPdel\tant\cost^2-\tanht\NPdel\tanht\cosht^2
-\frac{4\pi i\tant\cost^2}{3c}\J^{{4}},\\
\fl~\Upsilon^{{2}}_{{1}{}{1}}
&=&\Upsilon^{{1}}_{{2}{}{2}}
=\Upsilon^{{3}}_{{4}{}{4}}=0,\\
\fl {^\pm\!}\Upsilon^{{2}}_{{2}{}{3}}
&=&\mp\frac{1}{2}(\tau w +\pi^* w^*)
-\frac{1}{2}(\mp\NPdel\tanht\cosht^2-\tant\NPdel\tant\cost^2)\OPMIV
\pm\frac{2\pi(2\mp 3i\tant)}{3c\OMPIV}\J^{{4}}\\
\fl &=&\mp\frac{1}{2}(i\NPdel\tant\cost^2\OMPIV
-\NPdel\tanht\cosht^2\mp\tant\NPdel\tant\cost^2)\OPMIV
\mp\frac{2\pi}{3c\OMPIV}\J^{{4}}\\
\fl &=&\pm\frac{1}{2}(\NPdel\tanht\cosht^2-i\NPdel\tant\cost^2)\OPMIV
\mp\frac{2\pi}{3c\OMPIV}\J^{{4}},\\
\fl {^\pm\!}\Upsilon^{{1}}_{{1}{}{3}}
&=&\mp\frac{1}{2}(\tau w +\pi^* w^*)
-\frac{1}{2}(\pm\NPdel\tanht\cosht^2-\tant\NPdel\tant\cost^2)\OPMIV
\pm\frac{2\pi(2\mp 3i\tant)}{3c\OMPIV}\J^{{4}}\\
\fl &=&\mp\frac{1}{2}(i\NPdel\tant\cost^2\OMPIV
+\NPdel\tanht\cosht^2\mp\tant\NPdel\tant\cost^2)\OPMIV
\mp\frac{2\pi}{3c\OMPIV}\J^{{4}}\\
\fl &=&\mp\frac{1}{2}(\NPdel\tanht\cosht^2+i\NPdel\tant\cost^2)\OPMIV
\mp\frac{2\pi}{3c\OMPIV}\J^{{4}},\\
\fl {^\pm\!}\Upsilon^{{3}}_{{1}{}{3}}
&=&\pm\frac{1}{2}(\rho w +\rho^* w^*)
+\frac{1}{2}(\pm i\NPD\tant\cost^2-\tanht\NPD\tanht\cosht^2)\OMPU
\pm\frac{2\pi(2\pm 3\tanht)}{3c\OPMU}\J^{{2}}\\
\fl &=&\pm\frac{1}{2}(\NPD\tanht\cosht^2\OPMU
+i\NPD\tant\cost^2\mp\tanht\NPD\tanht\cosht^2)\OMPU
\mp\frac{2\pi}{3c\OPMU}\J^{{2}}\\
\fl &=&\pm\frac{1}{2}(\NPD\tanht\cosht^2+i\NPD\tant\cost^2)\OMPU
\mp\frac{2\pi}{3c\OPMU}\J^{{2}},\\
\fl {^\pm\!}\Upsilon^{{3}}_{{2}{}{3}}
&=&\pm\frac{1}{2}(\mu w +\mu^* w^*)
+\frac{1}{2}(\pm i\NPDEL\tant\cost^2-\tanht\NPDEL\tanht\cosht^2)\OPMU
\pm\frac{2\pi(2\mp 3\tanht)}{3c\OMPU}\J^{{1}}\\
\fl &=&\pm\frac{1}{2}(-\NPDEL\tanht\cosht^2\OMPU
+i\NPDEL\tant\cost^2\mp\tanht\NPDEL\tanht\cosht^2)\OPMU
\mp\frac{2\pi}{3c\OMPU}\J^{{1}}\\
\fl &=&\mp\frac{1}{2}(\NPDEL\tanht\cosht^2-i\NPDEL\tant\cost^2)\OPMU
\mp\frac{2\pi}{3c\OMPU}\J^{{1}},\\
\fl {^\pm\!}\Upsilon^{{4}}_{{1}{}{2}}
&=&\frac{1}{2\OPMU}\left(
\pm\tau w\OPMIV\mp\pi^* w^*\OMPIV
\pm\NPdel\tanht\frac{\cosht^2}{\cost^2}+\tant\NPdel\tant
-\frac{4\pi}{c}i\tant \J^{{4}}\right)\\
\fl &=&\frac{\pm 1}{2\OPMU}\left(
\NPdel\tanht\frac{\cosht^2}{\cost^2}
+\tau w-\pi^* w^*\right)\!,\\
\fl {^\pm\!}\Upsilon^{{2}}_{{3}{}{4}}
\fl &=&\frac{1}{2\OPMIV}\left(
\pm\rho w\OPMU\mp\rho^* w^*\OMPU
\pm i\NPD\tant\frac{\cost^2}{\cosht^2}-\tanht\NPD\tanht
+\frac{4\pi}{c}\tanht \J^{{2}}\right)\\
\fl &=&\frac{\pm 1}{2\OPMIV}\left(
i\NPD\tant\frac{\cost^2}{\cosht^2}
+\rho w-\rho^* w^*\right)\!,\\
\fl {^\pm\!}\Upsilon^{{1}}_{{4}{}{3}}
&=&\frac{1}{2\OMPIV}\left(
\pm\mu w\OMPU\mp\mu^* w^*\OPMU
\mp i\NPDEL\tant\frac{\cost^2}{\cosht^2}-\tanht\NPDEL\tanht
-\frac{4\pi}{c}\tanht \J^{{1}}\right)\\
\fl &=&\frac{\mp 1}{2\OMPIV}\left(
i\NPDEL\tant\frac{\cost^2}{\cosht^2}
-\mu w+\mu^* w^*\right)\!,\\
\fl {^\pm\!}\Upsilon^{{2}}_{{1}{}{3}}
&=&\frac{\kappa w(\tanht\!\mp\!1)}{\Zht}~~~,~~~
{^\pm\!}\Upsilon^{{1}}_{{2}{}{4}}
=-\frac{\nu w(\tanht\!\pm\!1)}{\Zht},\\
\fl {^\pm\!}\Upsilon^{{4}}_{{1}{}{3}}
&=&\frac{\sigma w(i\tant\!\pm\!1)}{\Zt}~~,~~~
{^\pm\!}\Upsilon^{{3}}_{{2}{}{4}}
=-\frac{\lambda w(i\tant\!\mp\!1)}{\Zt},\\
\fl ~~\Upsilon^{{4}}_{{1}{}{1}}
&=&\frac{\kappa w^2}{\Zht}~~~~~~~~~~,~~~~
\Upsilon^{{3}}_{{2}{}{2}}
=-\frac{\nu w^2}{\Zht},\\
\fl ~~\Upsilon^{{2}}_{{3}{}{3}}
&=&\frac{\sigma w^2}{\Zt}~~~~~~~~~~,~~~~
\Upsilon^{{1}}_{{4}{}{4}}
=-\frac{\lambda w^2}{\Zt}.
%\fl~~\Upsilon^{(\alpha)}_{\!(\beta)(\alpha)}
%\!&=&\Upsilon^{(\alpha)}_{\!((\beta)(\alpha))}
%=\frac{(\rmNt\,)_{,(\beta)}}{\rmNt}
%\!-\!\frac{4\pi}{3c}\frac{\rmg}{\rmN}
%\J^\rho \N_{[\rho\beta]}
%=\tant\tant_{,(\beta)}\cost^2-\tanht\tanht_{,(\beta)}\cosht^2,\\
%\fl~~\Upsilon^{(\alpha)}_{\![(\beta)(\alpha)]}\!&=&0.
\end{eqnarray}
%\bigskip\\
\section{\label{Check}Check of the Approximate Connection Addition Formula}
Here we will show that the order $f^2$ approximation of
$\Upsilon^{\alpha}_{\!\sigma\mu}$ in
(\ref{upsilonsymmetric},\ref{upsilonantisymmetric},\ref{upsiloncontracted})
matches the exact solution in \S\ref{Upsilon} for
$\cost\!=\!\cosht\!=\!\Zt\!=\!\Zht\!=\!1$,
which amounts to a second order approximation in $\tant~{\rm and}~\tanht$.
Much use is made of $g_{{a}{b}}$ and $f_{{a}{b}}$ from
(\ref{gtetrad},\ref{ftetrad}),
$\gamma_{{c}{}{a}{}{b}}=-\gamma_{{a}{}{c}{}{b}}$
from (\ref{gammasymmetry}),
$\ff_{,{a}}/4\!=\!\tanht\tanht_{,{a}}\!-\!\tant\tant_{,{a}}$
from (\ref{ff}), and the field equations
(\ref{Ampere1},\ref{Ampere2},\ref{Ampere3}).
To save space, only one component of each type will be shown.

In tetrad form (\ref{upsilonsymmetric}) becomes,
\begin{eqnarray}
\fl{\Upsilon}_{{c}\,({d}{}{e})}
&\approx&\frac{1}{2}
(f^{{a}}{_{{d}}}(f_{{e}{}{c},{a}}
-\gamma^{{b}}_{{e}{}{a}}f_{{b}{}{c}}
-\gamma^{{b}}_{{c}{}{a}}f_{{e}{}{b}})
+f^{{a}}{_{{e}}}(f_{{d}{}{c},{a}}
-\gamma^{{b}}_{{d}{}{a}}f_{{b}{}{c}}
-\gamma^{{b}}_{{c}{}{a}}f_{{d}{}{b}})\nonumber\\
\fl &&+f_{{c}}{^{{a}}}(f_{{a}{}{d},{e}}
-\gamma^{{b}}_{{a}{}{e}}f_{{b}{}{d}}
-\gamma^{{b}}_{{d}{}{e}}f_{{a}{}{b}})
+f_{{c}}{^{{a}}}(f_{{a}{}{e},{d}}
-\gamma^{{b}}_{{a}{}{d}}f_{{b}{}{e}}
-\gamma^{{b}}_{{e}{}{d}}f_{{a}{}{b}}))\nonumber\\
\fl &&+\frac{1}{8}(\ff_,{_{{c}}}g_{{d}{}{e}}
-\ff_{,{d}}g_{{e}{}{c}}
-\ff_{,{e}}g_{{d}{}{c}})\nonumber\\
\fl &&+\frac{2\pi}{c}\!\left(\J^{{a}}f_{{c}{a}}g_{{d}{e}}
+\frac{1}{3}\J^{{a}}f_{{a}{d}}g_{{e}{c}}
+\frac{1}{3}\J^{{a}}f_{{a}{e}}g_{{d}{c}}\right)\!,\\
\fl{\Upsilon}_{{1}{}({1}{}{2})}
&\approx&\frac{1}{2}
(f^{{a}}{_{{1}}}(f_{{2}{}{1},{a}}
-\gamma^{{b}}_{{2}{}{a}}f_{{b}{}{1}}
-\gamma^{{b}}_{{1}{}{a}}f_{{2}{}{b}})
+f^{{a}}{_{{2}}}(f_{{1}{}{1},{a}}
-\gamma^{{b}}_{{1}{}{a}}f_{{b}{}{1}}
-\gamma^{{b}}_{{1}{}{a}}f_{{1}{}{b}})\nonumber\\
\fl &&+f_{{1}}{^{{a}}}(f_{{a}{}{1},{2}}
-\gamma^{{b}}_{{a}{}{2}}f_{{b}{}{1}}
-\gamma^{{b}}_{{1}{}{2}}f_{{a}{}{b}})
+f_{{1}}{^{{a}}}(f_{{a}{}{2},{1}}
-\gamma^{{b}}_{{a}{}{1}}f_{{b}{}{2}}
-\gamma^{{b}}_{{2}{}{1}}f_{{a}{}{b}}))\nonumber\\
\fl &&+\frac{1}{8}(\ff_{,{1}} g_{{1}{}{2}}
-\ff_{,{1}} g_{{2}{}{1}}
-\ff_{,{2}} g_{{1}{}{1}})\nonumber\\
\fl &&+\frac{2\pi}{c}\!\left(\J^{{2}}f_{{1}{2}}g_{{1}{2}}
+\frac{1}{3}\J^{{2}}f_{{2}{1}}g_{{2}{1}}
+\frac{1}{3}\J^{{1}}f_{{1}{2}}g_{{1}{1}}\right)\\
\fl&=&\tanht\NPD\tanht-\frac{4\pi}{3c}\tanht \J^{{2}},\\
\fl{\Upsilon}_{{2}{}({1}{}{1})}
&\approx&
~~~f^{{a}}{_{{1}}}(f_{{1}{}{2},{a}}
-\gamma^{{b}}_{{1}{}{a}}f_{{b}{}{2}}
-\gamma^{{b}}_{{2}{}{a}}f_{{1}{}{b}})
+f_{{2}}{^{{a}}}(f_{{a}{}{1},{1}}
-\gamma^{{b}}_{{a}{}{1}}f_{{b}{}{1}}
-\gamma^{{b}}_{{1}{}{1}}f_{{a}{}{b}})\nonumber\\
\fl &&+\frac{1}{8}(\ff_{,{2}} g_{{1}{}{1}}
-\ff_{,{1}} g_{{1}{}{2}}
-\ff_{,{1}} g_{{1}{}{2}})\nonumber\\
\fl &&+\frac{2\pi}{c}\!\left(\J^{{1}}f_{{2}{1}}g_{{1}{1}}
+\frac{1}{3}\J^{{2}}f_{{2}{1}}g_{{1}{2}}
+\frac{1}{3}\J^{{2}}f_{{2}{1}}g_{{1}{2}}\right)\\
%\fl &=&-\frac{1}{4}\NPD\ff\\
\fl &=&\tant\NPD\tant\!-\!\tanht\NPD\tanht
+\frac{4\pi}{3c}\tanht \J^{{2}},\\
\fl{\Upsilon}_{{1}{}({1}{}{1})}
&\approx&
~~~f^{{a}}{_{{1}}}(f_{{1}{}{1},{a}}
-\gamma^{{b}}_{{1}{}{a}}f_{{b}{}{1}}
-\gamma^{{b}}_{{1}{}{a}}f_{{1}{}{b}})
+f_{{1}}{^{{a}}}(f_{{a}{}{1},{1}}
-\gamma^{{b}}_{{a}{}{1}}f_{{b}{}{1}}
-\gamma^{{b}}_{{1}{}{1}}f_{{a}{}{b}})\nonumber\\
\fl &&+\frac{1}{8}(\ff_{,{1}} g_{{1}{}{1}}
-\ff_{,{1}} g_{{1}{}{1}}
-\ff_{,{1}} g_{{1}{}{1}})\nonumber\\
\fl &&+\frac{2\pi}{c}\!\left(\J^{{2}}f_{{1}{2}}g_{{1}{1}}
+\frac{1}{3}\J^{{2}}f_{{2}{1}}g_{{1}{1}}
+\frac{1}{3}\J^{{2}}f_{{2}{1}}g_{{1}{1}}\right)\\
\fl &=&0,\\
\fl{\Upsilon}_{{1}{}({2}{}{3})}
&\approx&\frac{1}{2}
(f^{{a}}{_{{2}}}(f_{{3}{}{1},{a}}
-\gamma^{{b}}_{{3}{}{a}}f_{{b}{}{1}}
-\gamma^{{b}}_{{1}{}{a}}f_{{3}{}{b}})
+f^{{a}}{_{{3}}}(f_{{2}{}{1},{a}}
-\gamma^{{b}}_{{2}{}{a}}f_{{b}{}{1}}
-\gamma^{{b}}_{{1}{}{a}}f_{{2}{}{b}})\nonumber\\
\fl &&+f_{{1}}{^{{a}}}(f_{{a}{}{2},{3}}
-\gamma^{{b}}_{{a}{}{3}}f_{{b}{}{2}}
-\gamma^{{b}}_{{2}{}{3}}f_{{a}{}{b}})
+f_{{1}}{^{{a}}}(f_{{a}{}{3},{2}}
-\gamma^{{b}}_{{a}{}{2}}f_{{b}{}{3}}
-\gamma^{{b}}_{{3}{}{2}}f_{{a}{}{b}}))\nonumber\\
\fl &&+\frac{1}{8}(\ff_{,{1}} g_{{2}{}{3}}
-\ff_{,{2}} g_{{3}{}{1}}
-\ff_{,{3}} g_{{2}{}{1}})\nonumber\\
\fl &&+\frac{2\pi}{c}\!\left(\J^{{2}}f_{{1}{2}}g_{{2}{3}}
+\frac{1}{3}\J^{{1}}f_{{1}{2}}g_{{3}{1}}
+\frac{1}{3}\J^{{4}}f_{{4}{3}}g_{{2}{1}}\right)\\
\fl&=&\frac{1}{2}(i\tant\NPdel\tanht+\tanht\NPdel\tanht)
-\frac{1}{2}(\tanht\NPdel\tanht\!-\!\tant\NPdel\tant)
-\frac{2\pi}{3c}i\tant \J^{{4}}\\
\fl&=&\frac{i\tant}{2}(\NPdel\tanht-i\NPdel\tant)
-\frac{2\pi}{3c}i\tant \J^{{4}},\\
\fl{\Upsilon}_{{3}{}({1}{}{2})}
&\approx&\frac{1}{2}
(f^{{a}}{_{{1}}}(f_{{2}{}{3},{a}}
-\gamma^{{b}}_{{2}{}{a}}f_{{b}{}{3}}
-\gamma^{{b}}_{{3}{}{a}}f_{{2}{}{b}})
+f^{{a}}{_{{2}}}(f_{{1}{}{3},{a}}
-\gamma^{{b}}_{{1}{}{a}}f_{{b}{}{3}}
-\gamma^{{b}}_{{3}{}{a}}f_{{1}{}{b}})\nonumber\\
\fl &&+f_{{3}}{^{{a}}}(f_{{a}{}{1},{2}}
-\gamma^{{b}}_{{a}{}{2}}f_{{b}{}{1}}
-\gamma^{{b}}_{{1}{}{2}}f_{{a}{}{b}})
+f_{{3}}{^{{a}}}(f_{{a}{}{2},{1}}
-\gamma^{{b}}_{{a}{}{1}}f_{{b}{}{2}}
-\gamma^{{b}}_{{2}{}{1}}f_{{a}{}{b}}))\nonumber\\
\fl &&+\frac{1}{8}(\ff_{,{3}} g_{{1}{}{2}}
-\ff_{,{1}} g_{{2}{}{3}}
-\ff_{,{2}} g_{{1}{}{3}})\nonumber\\
\fl &&+\frac{2\pi}{c}\!\left(\J^{{4}}f_{{3}{4}}g_{{1}{2}}
+\frac{1}{3}\J^{{2}}f_{{2}{1}}g_{{2}{3}}
+\frac{1}{3}\J^{{1}}f_{{1}{2}}g_{{1}{3}}\right)\\
\fl &=& \frac{1}{2}
\tanht(-\gamma_{{3}{}{2}{}{1}}i\tant-\gamma_{{2}{}{3}{}{1}}\tanht
+\gamma_{{3}{}{1}{}{2}}i\tant-\gamma_{{1}{}{3}{}{2}}\tanht)\nonumber\\
\fl&+&\frac{1}{2}
i\tant(\gamma_{{1}{}{3}{}{2}}\tanht-\gamma_{{3}{}{1}{}{2}}i\tant
-\gamma_{{2}{}{3}{}{1}}\tanht-\gamma_{{3}{}{2}{}{1}}i\tant)
+\frac{1}{2}(\tanht\NPdel\tanht\!-\!\tant\NPdel\tant)
+\frac{2\pi}{c}i\tant \J^{{4}}\\
\fl &=& \frac{1}{2}\tanht(\NPdel\tanht-\pi^*\tanht+\tau\tanht)
+\frac{1}{2}i\tant(i\NPdel\tant-\tau i\tant+\pi^*i\tant)
+\frac{2\pi}{c}i\tant \J^{{4}}\\
\fl &=& \frac{1}{2}\tanht(\NPdel\tanht-\pi^*\tanht+\tau\tanht)
+\frac{1}{2}i\tant(\tau w+\pi^*w^*-\tau i\tant+\pi^*i\tant)\\
\fl &=& \frac{1}{2}\tanht(\NPdel\tanht-\pi^*\tanht+\tau\tanht)
+\frac{1}{2}i\tant(\tau \tanht+\pi^*\tanht)\\
\fl &=& \frac{1}{2}\tanht(\NPdel\tanht-\pi^*\tanht+\tau\tanht
+\tau i\tant+\pi^*i\tant)\\
\fl &=& \frac{1}{2}\tanht(\NPdel\tanht+\tau w-\pi^*w^*),\\
\fl{\Upsilon}_{{1}{}({1}{}{3})}
&\approx&\frac{1}{2}
(f^{{a}}{_{{1}}}(f_{{3}{}{1},{a}}
-\gamma^{{b}}_{{3}{}{a}}f_{{b}{}{1}}
-\gamma^{{b}}_{{1}{}{a}}f_{{3}{}{b}})
+f^{{a}}{_{{3}}}(f_{{1}{}{1},{a}}
-\gamma^{{b}}_{{1}{}{a}}f_{{b}{}{1}}
-\gamma^{{b}}_{{1}{}{a}}f_{{1}{}{b}})\nonumber\\
\fl &&+f_{{1}}{^{{a}}}(f_{{a}{}{1},{3}}
-\gamma^{{b}}_{{a}{}{3}}f_{{b}{}{1}}
-\gamma^{{b}}_{{1}{}{3}}f_{{a}{}{b}})
+f_{{1}}{^{{a}}}(f_{{a}{}{3},{1}}
-\gamma^{{b}}_{{a}{}{1}}f_{{b}{}{3}}
-\gamma^{{b}}_{{3}{}{1}}f_{{a}{}{b}}))\nonumber\\
\fl &&+\frac{1}{8}(\ff_{,{1}} g_{{1}{}{3}}
-\ff_{,{1}} g_{{3}{}{1}}
-\ff_{,{3}} g_{{1}{}{1}})\nonumber\\
\fl &&+\frac{2\pi}{c}\!\left(\J^{{2}}f_{{1}{2}}g_{{1}{3}}
+\frac{1}{3}\J^{{2}}f_{{2}{1}}g_{{3}{1}}
+\frac{1}{3}\J^{{4}}f_{{4}{3}}g_{{1}{1}}\right)\\
\fl &=&\tanht(-\gamma_{{1}{}{3}{}{1}}\tanht+\gamma_{{3}{}{1}{}{1}}i\tant)\\
\fl &=&\kappa\tanht w,\\
\fl{\Upsilon}_{{3}{}({1}{}{1})}
&\approx&
~~~f^{{a}}{_{{1}}}(f_{{1}{}{3},{a}}
-\gamma^{{b}}_{{1}{}{a}}f_{{b}{}{3}}
-\gamma^{{b}}_{{3}{}{a}}f_{{1}{}{b}})
+f_{{3}}{^{{a}}}(f_{{a}{}{1},{1}}
-\gamma^{{b}}_{{a}{}{1}}f_{{b}{}{1}}
-\gamma^{{b}}_{{1}{}{1}}f_{{a}{}{b}})\nonumber\\
\fl &&+\frac{1}{8}(\ff_{,{3}} g_{{1}{}{1}}
-\ff_{,{1}} g_{{1}{}{3}}
-\ff_{,{1}} g_{{1}{}{3}})\nonumber\\
\fl &&+\frac{2\pi}{c}\!\left(\J^{{4}}f_{{3}{4}}g_{{1}{1}}
+\frac{1}{3}\J^{{2}}f_{{2}{1}}g_{{1}{3}}
+\frac{1}{3}\J^{{2}}f_{{2}{1}}g_{{1}{3}}\right)\\
\fl &=&\tanht(-\gamma_{{3}{}{1}{}{1}}i\tant+\gamma_{{1}{}{3}{}{1}}\tanht)
-i\tant(-\gamma_{{1}{}{3}{}{1}}\tanht+\gamma_{{3}{}{1}{}{1}}i\tant)\\
\fl &=&-\kappa w^2.
\end{eqnarray}
%\pagebreak
%\bigskip\\
%\bigskip\\
In tetrad form (\ref{upsilonantisymmetric}) becomes,
\begin{eqnarray}
\fl{\Upsilon}_{{c}\,[{d}{}{e}]}
&\approx&\frac{1}{2}(f_{{d}{}{e},}{_{{c}}}
\!-\!\gamma^{{b}}_{{d}{}{c}}f_{{b}{}{e}}
\!-\!\gamma^{{b}}_{{e}{}{c}}f_{{d}{}{b}}
+f_{{c}{}{e},}{_{{d}}}
\!-\!\gamma^{{b}}_{{c}{}{d}}f_{{b}{}{e}}
\!-\!\gamma^{{b}}_{{e}{}{d}}f_{{c}{}{b}}
-f_{{c}{}{d},}{_{{e}}}
\!+\!\gamma^{{b}}_{{c}{}{e}}f_{{b}{}{d}}
\!+\!\gamma^{{b}}_{{d}{}{e}}f_{{c}{}{b}})\nonumber\\
\fl &&+\frac{4\pi}{3c}\left(\J_{{d}}g_{{e}{c}}
-\J_{{e}}g_{{d}{c}}\right),\\
\fl{\Upsilon}_{{1}{}[{1}{}{2}]}
&\approx&\frac{1}{2}(f_{{1}{}{2},}{_{{1}}}
\!-\!\gamma^{{b}}_{{1}{}{1}}f_{{b}{}{2}}
\!-\!\gamma^{{b}}_{{2}{}{1}}f_{{1}{}{b}}
+f_{{1}{}{2},}{_{{1}}}
\!-\!\gamma^{{b}}_{{1}{}{1}}f_{{b}{}{2}}
\!-\!\gamma^{{b}}_{{2}{}{1}}f_{{1}{}{b}}
-f_{{1}{}{1},}{_{{2}}}
\!+\!\gamma^{{b}}_{{1}{}{2}}f_{{b}{}{1}}
\!+\!\gamma^{{b}}_{{1}{}{2}}f_{{1}{}{b}})\nonumber\\
\fl &&+\frac{4\pi}{3c}\left(\J_{{1}}g_{{2}{1}}
-\J_{{2}}g_{{1}{1}}\right)\\
\fl &=&-\NPD\tanht
+\frac{4\pi}{3c}\J^{{2}},\\
\fl{\Upsilon}_{{1}{}[{2}{}{3}]}
&\approx&\frac{1}{2}(f_{{2}{}{3},}{_{{1}}}
\!-\!\gamma^{{b}}_{{2}{}{1}}f_{{b}{}{3}}
\!-\!\gamma^{{b}}_{{3}{}{1}}f_{{2}{}{b}}
+f_{{1}{}{3},}{_{{2}}}
\!-\!\gamma^{{b}}_{{1}{}{2}}f_{{b}{}{3}}
\!-\!\gamma^{{b}}_{{3}{}{2}}f_{{1}{}{b}}
-f_{{1}{}{2},}{_{{3}}}
\!+\!\gamma^{{b}}_{{1}{}{3}}f_{{b}{}{2}}
\!+\!\gamma^{{b}}_{{2}{}{3}}f_{{1}{}{b}})\nonumber\\
\fl &&+\frac{4\pi}{3c}\left(\J_{{2}}g_{{3}{1}}
-\J_{{3}}g_{{2}{1}}\right)\\
\fl &=&\frac{1}{2}(\NPdel\tanht
-\gamma_{{3}{}{2}{}{1}}i\tant-\gamma_{{2}{}{3}{}{1}}\tanht
-\gamma_{{3}{}{1}{}{2}}i\tant+\gamma_{{1}{}{3}{}{2}}\tanht)
-\frac{4\pi}{3c}\J_{{3}}\\
\fl &=&\frac{1}{2}(\NPdel\tanht-\pi^* w^*-\tau w)
-\frac{4\pi}{3c}\J_{{3}}\\
\fl &=&\frac{1}{2}(\NPdel\tanht-i\NPdel\tant)
-\frac{2\pi}{3c}\J^{{4}},\\
\fl{\Upsilon}_{{3}{}[{1}{}{2}]}
&\approx&\frac{1}{2}(f_{{1}{}{2},}{_{{3}}}
\!-\!\gamma^{{b}}_{{1}{}{3}}f_{{b}{}{2}}
\!-\!\gamma^{{b}}_{{2}{}{3}}f_{{1}{}{b}}
+f_{{3}{}{2},}{_{{1}}}
\!-\!\gamma^{{b}}_{{3}{}{1}}f_{{b}{}{2}}
\!-\!\gamma^{{b}}_{{2}{}{1}}f_{{3}{}{b}}
-f_{{3}{}{1},}{_{{2}}}
\!+\!\gamma^{{b}}_{{3}{}{2}}f_{{b}{}{1}}
\!+\!\gamma^{{b}}_{{1}{}{2}}f_{{3}{}{b}})\nonumber\\
\fl &&+\frac{4\pi}{3c}\left(\J_{{1}}g_{{2}{3}}
-\J_{{2}}g_{{1}{3}}\right)\\
\fl &=&\frac{1}{2}(-\NPdel\tanht
+\gamma_{{2}{}{3}{}{1}}\tanht+\gamma_{{3}{}{2}{}{1}}i\tant
+\gamma_{{1}{}{3}{}{2}}\tanht-\gamma_{{3}{}{1}{}{2}}i\tant)\\
\fl &=&-\frac{1}{2}(\NPdel\tanht+\tau w-\pi^* w^*),\\
\fl{\Upsilon}_{{1}{}[{1}{}{3}]}
&\approx&\frac{1}{2}(f_{{1}{}{3},}{_{{1}}}
\!-\!\gamma^{{b}}_{{1}{}{1}}f_{{b}{}{3}}
\!-\!\gamma^{{b}}_{{3}{}{1}}f_{{1}{}{b}}
+f_{{1}{}{3},}{_{{1}}}
\!-\!\gamma^{{b}}_{{1}{}{1}}f_{{b}{}{3}}
\!-\!\gamma^{{b}}_{{3}{}{1}}f_{{1}{}{b}}
-f_{{1}{}{1},}{_{{3}}}
\!+\!\gamma^{{b}}_{{1}{}{3}}f_{{b}{}{1}}
\!+\!\gamma^{{b}}_{{1}{}{3}}f_{{1}{}{b}})\nonumber\\
\fl &&+\frac{4\pi}{3c}\left(\J_{{1}}g_{{3}{1}}
-\J_{{3}}g_{{1}{1}}\right)\\
\fl &=&-\gamma_{{3}{}{1}{}{1}}i\tant+\gamma_{{1}{}{3}{}{1}}\tanht\\
\fl &=&-\kappa w.
\end{eqnarray}

\section{\label{Proof}The Non-symmetric Matrix Decomposition Theorem}

$\mathbf{Theorem}$: Assume $\Nbar^{\sigma\mu}$ is a real tensor,
$f^{\sigma\mu}\!=\!{\Nbar}^{[\sigma\mu]}$, and
$g^{\sigma\mu}\!=\!{\Nbar}^{(\sigma\mu)}$ is an invertible metric
which can be put into Newman-Penrose tetrad form
\begin{eqnarray}
\label{gtetrad2}
~g_{{a}{}{b}}
&=&~g^{{a}{}{b}}
=g^{\alpha\beta}e^{{a}}{_{\alpha}}\,e^{{b}}{_{\beta}}
=\pmatrix{
~0&1&0&0\cr
~1&0&0&0\cr
~0&0&0&\!\!\!-1\cr
~0&0&\!\!-1&0
},\\
%\end{eqnarray}
%\begin{eqnarray}
~~l^\sigma\!&=&e_{{1}}{^\sigma}\,,\,
n^\sigma\!=e_{{2}}{^\sigma}\,,\,
m^\sigma\!=e_{{3}}{^\sigma}~~,\,
m^{\!*\,\sigma}\!=e_{{4}}{^\sigma},\\
~~l_\sigma\!&=&e^{{2}}{_\sigma}\,,\,
n_\sigma\!=e^{{1}}{_\sigma}\,,\,
m_\sigma\!=-e^{{4}}{_\sigma}\,,\,
m^*_\sigma\!=-e^{{3}}{_\sigma},\\
~\delta^\sigma_\mu
\!&=&e_{{a}}{^\sigma}e^{{a}}{_\mu}~~~~~~~~~,~
\delta^{{a}}_{{b}}=e_{{b}}{^\sigma}e^{{a}}{_\sigma},
\end{eqnarray}
where $l_\sigma$ and $n_\sigma$ are real, $m_\sigma$ and $m^*_\sigma$ are
complex conjugates. Then tetrads may be chosen such that
\begin{eqnarray}
\label{decomposition1}
\fl~~~~\Nbar^{{a}{}{b}}
&=&\Nbar^{\alpha\beta}e^{{a}}{_{\alpha}}\,e^{{b}}{_{\beta}}
=\pmatrix{
0&\!\!(1\!+\!\tanht)&0&0\cr
(1\!-\!\tanht)&0&0&0\cr
0&0&0&\!\!-(1\!+\!i\tant)\cr
0&0&\!\!-(1\!-\!i\tant)&0
},
\end{eqnarray}
where $\tant$ and $\tanht$ are real, except
if $f^\sigma{_\mu}f^\mu{_\sigma}\!=det(f^{\mu}{_{\nu}})\!=0$, in which case
tetrads may be chosen such that
\begin{eqnarray}
\label{decomposition2}
\fl~~~~\Nbar^{{a}{}{b}}
&=&\Nbar^{\alpha\beta}e^{{a}}{_{\alpha}}\,e^{{b}}{_{\beta}}
=\pmatrix{
0&1&0&0\cr
1&0&\!\!\!-\uacute&\!\!\!-\uacute\cr
0&\uacute&0&\!\!\!-1\cr
0&\uacute&\!\!\!-1&0
},
\end{eqnarray}
where $\uacute$ is real.

$\mathbf{Proof}$: From \cite{Chandrasekhar} p.51, $f_{{a}{}{b}}$ can
be parameterized by the three complex scalars
\begin{eqnarray}
\label{complexscalars}
\fl~~~~~~~\phi_0=f_{{1}{}{3}}~~~,~~~
\phi_1=(f_{{1}{}{2}}+f_{{4}{}{3}})/2~~~,~~~
\phi_2=f_{{4}{}{2}}.
\end{eqnarray}
From \cite{Chandrasekhar} p.53-54, there are a series of tetrad transformations
which do not alter (\ref{gtetrad2}).
\noindent
Type I:
\begin{eqnarray}
\fl l_\sigma\rightarrow l_\sigma,~~
m_\sigma\rightarrow m_\sigma\!+\!al_\sigma,~~
m_\sigma^\star\rightarrow m_\sigma^\star\!+\!a^\star l_\sigma,~~
n_\sigma\rightarrow
n_\sigma\!+\!a^\star m_\sigma\!+\!a m_\sigma^\star\!+\!aa^\star l_\sigma,\\
\fl \phi_0\rightarrow\phi_0,~~
\phi_1\rightarrow\phi_1\!+\!a^\star\phi_0,~~
\phi_2\rightarrow\phi_2\!+\!2a^\star\phi_1\!+\!(a^\star)^2\phi_0.
\end{eqnarray}
Type II:
\begin{eqnarray}
\fl n_\sigma\rightarrow n_\sigma,~~
m_\sigma\rightarrow m_\sigma\!+\!bn_\sigma,~~
m_\sigma^\star\rightarrow m_\sigma^\star\!+\!b^\star n_\sigma,~~
l_\sigma\rightarrow
l_\sigma\!+\!b^\star m_\sigma\!+\!b m_\sigma^\star\!+\!bb^\star n_\sigma,\\
\fl \phi_2\rightarrow\phi_2,~~
\phi_1\rightarrow\phi_1\!+\!b\phi_2,~~
\phi_0\rightarrow\phi_0\!+\!2b\phi_1\!+\!b^2\phi_2.
\end{eqnarray}
Type III:
\begin{eqnarray}
\fl n_\sigma\rightarrow n_\sigma A,~~
l_\sigma\rightarrow l_\sigma/A,~~
m_\sigma\rightarrow m_\sigma e^{i\theta},~~
m_\sigma^\star\rightarrow m_\sigma^\star e^{-i\theta},\\
\fl \phi_0\rightarrow\phi_0 e^{i\theta}/A ,~~
\phi_1\rightarrow\phi_1,~~
\phi_2\rightarrow\phi_0 e^{-i\theta}A.
\end{eqnarray}

Using type I and II transformations, we can always make
either $\phi_2\!=\!0$ or $\phi_0\!=\!0$ by solving a quadradic equation
and performing a tetrad transformation with
\begin{eqnarray}
\fl~~~a^\star=\frac{-2\phi_1\pm\sqrt{(2\phi_1)^2-4\phi_0\phi_2}}{2\phi_0}
~~~~{\rm or}~~~~
b~=\frac{-2\phi_1\pm\sqrt{(2\phi_1)^2-4\phi_2\phi_0}}{2\phi_2}.
\end{eqnarray}
Note that a type I transformation does not alter $\phi_0$ and a type II
transformation does not alter $\phi_2$. Therefore, if $\phi_1\!\ne\!0$ at this
point, we can make $\phi_0\!=\!\phi_2\!=\!0$ by doing a second transformation
of the opposite type to the first one with
\begin{eqnarray}
\label{secondtransformation}
\fl~~~b=-\frac{\phi_0}{2\phi_1}
~~~~{\rm or}~~~~
a^\star=-\frac{\phi_2}{2\phi_1}.
\end{eqnarray}
Then with
$\tanht\!=\!-2Re(\phi_1)$,
 $\tant\!=\!-2Im(\phi_1)$,
we have from (\ref{complexscalars},\ref{gtetrad2}),
\begin{eqnarray}
\label{mixed}
\fl f_{ab}\!=\!\pmatrix{
0&\!\!\!-\tanht&\!0&\!0\cr
\tanht&\!0&\!0&\!0\cr
0&\!0&\!0&\!i\tant\cr
0&\!0&\!\!\!-i\tant&\!0
}\!,
~f^a{_b}\!=\!\pmatrix{
\tanht&\!0&\!0&\!0\cr
0&\!\!\!-\tanht&\!0&\!0\cr
0&\!0&\!i\tant&\!0\cr
0&\!0&\!0&\!\!\!-i\tant
}\!,
~f^{ab}\!=\!\pmatrix{
0&\!\tanht&0&\!0\!\cr
\!\!-\tanht&\!0&0&\!0\!\cr
0&\!0&\!0&\!\!\!-i\tant\!\cr
0&\!0&\!i\tant&\!0\!
}\!.
\end{eqnarray}
This is the first case (\ref{decomposition1}). The procedure above fails if
$\phi_1\!=\!0$ in \ref{secondtransformation}, in which case there is only one
nonzero scalar, either $\phi_0$ or $\phi_2$. If the nonzero scalar is $\phi_2$,
it can be changed to $\phi_0$ by doing type II transformation with $b\!=\!1$
followed by a type I transformation with $a^*\!=\!-1$. Furthermore, we can make
$\phi_0$ real by doing a type III transformation. Then with $\uacute=\phi_0$
we have from (\ref{complexscalars},\ref{gtetrad2}),
\begin{eqnarray}
\label{mixed2}
\fl f_{{a}{}{b}}
\!=\!\pmatrix{
0&\!0&\uacute&\uacute\cr
0&\!0&0&\!0\cr
-\uacute&\!0&0&\!0\cr
-\uacute&\!0&0&\!0
}\!,~~~
f^{{a}}{_{{b}}}
\!=\!\pmatrix{
0&\!0&0&\!0\cr
0&\!0&\uacute&\uacute\cr
\uacute&\!0&0&\!0\cr
\uacute&\!0&0&\!0
}\!,~~~
f^{{a}{b}}
\!=\!\pmatrix{
0&0&\!0&0\!\cr
0&0&\!\!\!-\uacute&\!\!\!-\uacute\cr
0&\uacute&\!0&0\!\cr
0&\uacute&\!0&0\!
}\!.
%\label{contradiction}
%$f^{{a}}{}{_{{b}}}f^{{b}}{}{_{{a}}}
%\!=\!f^\sigma{_\mu}f^\mu{_\sigma}$
%$det(f^{{a}}{}{_{{b}}})\!=det(f^\mu{_{\nu}})$
\end{eqnarray}
This is the second case (\ref{decomposition2}). Since
$f^\sigma{_\mu}f^\mu{_\sigma}\!=\!f^{{a}}{}{_{{b}}}f^{{b}}{}{_{{a}}}$ and
$det(f^\mu{_{\nu}})\!=det(f^{{a}}{}{_{{b}}})$, we see from
(\ref{mixed},\ref{mixed2}) that this second case occurs if and only if
$f^\sigma{_\mu}f^\mu{_\sigma}\!=det(f^{\mu}{_{\nu}})\!=0$.
This proves the theorem.

%\pagebreak
%\bibliography{npshifflett}% Produces the bibliography via BibTeX.
%\section*{References}
%\begin{thebibliography}{68}
\Bibliography{68}
\expandafter\ifx\csname natexlab\endcsname\relax\def\natexlab#1{#1}\fi
\expandafter\ifx\csname bibnamefont\endcsname\relax
  \def\bibnamefont#1{#1}\fi
\expandafter\ifx\csname bibfnamefont\endcsname\relax
  \def\bibfnamefont#1{#1}\fi
\expandafter\ifx\csname citenamefont\endcsname\relax
  \def\citenamefont#1{#1}\fi
\expandafter\ifx\csname url\endcsname\relax
  \def\url#1{\texttt{#1}}\fi
\expandafter\ifx\csname urlprefix\endcsname\relax\def\urlprefix{URL }\fi
\providecommand{\bibinfo}[2]{#2}
\providecommand{\eprint}[2][]{#2}
\providecommand{\bitem}[2][]{\bibitem {#2}}

\bitem[{\citenamefont{Einstein and Straus}(1946)}]{EinsteinStraus}
\bibinfo{author}{\bibfnamefont{A.}~\bibnamefont{Einstein}} \bibnamefont{and}
  \bibinfo{author}{\bibfnamefont{E.~G.} \bibnamefont{Straus}},
  \bibinfo{journal}{Ann.\ Math.} \textbf{\bibinfo{volume}{47}}
  (\bibinfo{year}{1946}) \bibinfo{pages}{731}.

\bitem[{\citenamefont{Einstein}(1948)}]{Einstein3}
\bibinfo{author}{\bibfnamefont{A.}~\bibnamefont{Einstein}},
  \bibinfo{journal}{Rev.\ Mod.\ Phys.} \textbf{\bibinfo{volume}{20}}
  (\bibinfo{year}{1948}) \bibinfo{pages}{35}.

\bitem[{\citenamefont{Einstein}(1950)}]{EinsteinBianchi}
\bibinfo{author}{\bibfnamefont{A.}~\bibnamefont{Einstein}},
  \bibinfo{journal}{Can.\ J.\ Math.} \textbf{\bibinfo{volume}{2}}
  (\bibinfo{year}{1949}) \bibinfo{pages}{120}.

\bitem[{\citenamefont{Einstein and Kaufman}(1955)}]{EinsteinKaufman}
\bibinfo{author}{\bibfnamefont{A.}~\bibnamefont{Einstein}} \bibnamefont{and}
  \bibinfo{author}{\bibfnamefont{B.}~\bibnamefont{Kaufman}},
  \bibinfo{journal}{Ann.\ Math.} \textbf{\bibinfo{volume}{62}}
  (\bibinfo{year}{1955}) \bibinfo{pages}{128}.

\bitem[{\citenamefont{Einstein}(1956)}]{EinsteinMOR}
\bibinfo{author}{\bibfnamefont{A.}~\bibnamefont{Einstein}},
  \emph{\bibinfo{title}{The Meaning of Relativity, 5th edition, revised}}
  (\bibinfo{publisher}{Princeton University Press}, \bibinfo{address}{Princeton
  NJ}, \bibinfo{year}{1956}).

\bitem[{\citenamefont{Schr\"{o}dinger}(1947)}]{SchrodingerI}
\bibinfo{author}{\bibfnamefont{E.}~\bibnamefont{Schr\"{o}dinger}},
  \bibinfo{journal}{Proc.\ Royal Irish Acad.} \textbf{\bibinfo{volume}{51A}}
  (\bibinfo{year}{1947}) \bibinfo{pages}{163}.

\bitem[{\citenamefont{Schr\"{o}dinger}(1948)}]{SchrodingerII}
\bibinfo{author}{\bibfnamefont{E.}~\bibnamefont{Schr\"{o}dinger}},
  \bibinfo{journal}{Proc.\ Royal Irish Acad.} \textbf{\bibinfo{volume}{51A}}
  (\bibinfo{year}{1948}) \bibinfo{pages}{205}.

\bitem[{\citenamefont{Schr\"{o}dinger}(1948)}]{SchrodingerIII}
\bibinfo{author}{\bibfnamefont{E.}~\bibnamefont{Schr\"{o}dinger}},
  \bibinfo{journal}{Proc.\ Royal Irish Acad.} \textbf{\bibinfo{volume}{52A}}
  (\bibinfo{year}{1948}) \bibinfo{pages}{1}.

\bitem[{\citenamefont{Schr\"{o}dinger}(1950)}]{SchrodingerSTS}
\bibinfo{author}{\bibfnamefont{E.}~\bibnamefont{Schr\"{o}dinger}},
  \emph{\bibinfo{title}{Space-Time Structure}}
  (\bibinfo{publisher}{Cambridge Press}, \bibinfo{address}{London},
  \bibinfo{year}{1950}) p 93,108,112.

\bitem[{\citenamefont{Shifflett}(2003)}]{Shifflett}
\bibinfo{author}{\bibfnamefont{J.~A.} \bibnamefont{Shifflett}},
 \eprint{arXiv:gr-qc/0310124}.

\bitem[{\citenamefont{Shifflett}()}]{Shifflett3}
\bibinfo{author}{\bibfnamefont{J.~A.} \bibnamefont{Shifflett}},
  \eprint{arXiv:gr-qc/0411016}.

\bitem[{\citenamefont{Sahni and Starobinsky}(2000)}]{Sahni}
\bibinfo{author}{\bibfnamefont{V.}~\bibnamefont{Sahni}} \bibnamefont{and}
  \bibinfo{author}{\bibfnamefont{A.}~\bibnamefont{Starobinsky}},
  \bibinfo{journal}{Int.\ J.\ Mod.\ Phys.} \textbf{\bibinfo{volume}{D9}}
  (\bibinfo{year}{2000}) \bibinfo{pages}{373} [arXiv:astro-ph/9904398].

\bitem[{\citenamefont{Peskin}(1964)}]{Peskin}
\bibinfo{author}{\bibfnamefont{M.E.}~\bibnamefont{Peskin}},
\bibinfo{author}{\bibfnamefont{D.V.}~\bibnamefont{Schroeder}},
  \emph{\bibinfo{title}{An Introduction to Quantum Field Theory}}
  (\bibinfo{publisher}{Westview Press},
  \bibinfo{year}{1995}) p 790-791,402.

\bitem[{\citenamefont{Zeldovich}(1968)}]{Zeldovich}
\bibinfo{author}{\bibfnamefont{Ya.B.}~\bibnamefont{Zeldovich}},
  \bibinfo{journal}{Sov.\ Phys. - Uspekhi} \textbf{\bibinfo{volume}{11}}
  (\bibinfo{year}{1968}) \bibinfo{pages}{381}.

\bitem[{\citenamefont{Borchsenius}(1978)}]{Borchsenius}
\bibinfo{author}{\bibfnamefont{K.}~\bibnamefont{Borchsenius}},
  \bibinfo{journal}{Nuovo Cimento} \textbf{\bibinfo{volume}{46A}}
  (\bibinfo{year}{1978}) \bibinfo{pages}{403}.

\bitem[{\citenamefont{Antoci}(1991)}]{Antoci3}
\bibinfo{author}{\bibfnamefont{S.}~\bibnamefont{Antoci}},
  \bibinfo{journal}{Gen.\ Relativ.\ Gravit.} \textbf{\bibinfo{volume}{23}}
  (\bibinfo{year}{1991}) \bibinfo{pages}{47}  [arXiv:gr-qc/0108052].

\bitem[{\citenamefont{Adler et~al.}(1975)\citenamefont{Adler, Bazin, and
  Schiffer}}]{Adler}
\bibinfo{author}{\bibfnamefont{R.}~\bibnamefont{Adler}},
  \bibinfo{author}{\bibfnamefont{M.}~\bibnamefont{Bazin}}, \bibnamefont{and}
  \bibinfo{author}{\bibfnamefont{M.}~\bibnamefont{Schiffer}},
  \emph{\bibinfo{title}{Introduction to General Relativity }}
  (\bibinfo{publisher}{McGraw-Hill}, \bibinfo{address}{New York},
  \bibinfo{year}{1975}) p 87.

\bitem[{\citenamefont{H\'{e}ly}(1954)}]{Hely}
\bibinfo{author}{\bibfnamefont{J.}~\bibnamefont{H\'{e}ly}},
  \bibinfo{journal}{C.\ R.\ Acad.\ Sci.\ (Paris)}
  \textbf{\bibinfo{volume}{239}}
  (\bibinfo{year}{1954}) \bibinfo{pages}{385}.

\bitem[{\citenamefont{Treder}(1957)}]{Treder57}
\bibinfo{author}{\bibfnamefont{H.-J.}~\bibnamefont{Treder}},
  \bibinfo{journal}{Annalen der Physik} \textbf{\bibinfo{volume}{19}}
  (\bibinfo{year}{1957}) \bibinfo{pages}{369}.

\bitem[{\citenamefont{Johnson}(1985)}]{JohnsonI}
\bibinfo{author}{\bibfnamefont{C.~R.} \bibnamefont{Johnson}},
  \bibinfo{journal}{Phys.\ Rev.\ D} \textbf{\bibinfo{volume}{31}}
  (\bibinfo{year}{1985}) \bibinfo{pages}{1236}.

\bitem[{\citenamefont{Tonnelat}(1954)}]{Tonnelat}
\bibinfo{author}{\bibfnamefont{M.~A.} \bibnamefont{Tonnelat}},
  \bibinfo{journal}{C.\ R.\ Acad.\ Sci.} \textbf{\bibinfo{volume}{239}}
  (\bibinfo{year}{1954}) \bibinfo{pages}{231}.

\bitem[{\citenamefont{Hlavaty}(1957)}]{Hlavaty}
\bibinfo{author}{\bibfnamefont{V.}~\bibnamefont{Hlavaty}},
  \emph{\bibinfo{title}{Geometry of Einstein's Unified Field Theory}}
  (\bibinfo{publisher}{P. Noordhoff Ltd.},
  \bibinfo{address}{Groningen},
  \bibinfo{year}{1957}) p 11-16.

\bitem[{\citenamefont{Newman and Penrose}(1962)}]{Newman}
\bibinfo{author}{\bibfnamefont{E.~T.} \bibnamefont{Newman}} \bibnamefont{and}
  \bibinfo{author}{\bibfnamefont{R.}~\bibnamefont{Penrose}},
  \bibinfo{journal}{J.\ Math.\ Phys.} \textbf{\bibinfo{volume}{3}}
  (\bibinfo{year}{1962}) \bibinfo{pages}{566}.

\bitem[{\citenamefont{Chandrasekhar}(1992)}]{Chandrasekhar}
\bibinfo{author}{\bibfnamefont{S.}~\bibnamefont{Chandrasekhar}},
  \emph{\bibinfo{title}{The Mathematical Theory of Black Holes}}
  (\bibinfo{publisher}{Oxford University Press},
  \bibinfo{address}{New York}, \bibinfo{year}{1992}) p 34,40,224,543.

\bitem[{\citenamefont{Padmanabhan2}(2002)}]{Padmanabhan2}
\bibinfo{author}{\bibfnamefont{T.} \bibnamefont{Padmanabhan}},
  \bibinfo{journal}{Class.\ Quantum\ Grav.} \textbf{\bibinfo{volume}{19}}
  (\bibinfo{year}{2002}) \bibinfo{pages}{3551}.

\bitem[{\citenamefont{Padmanabhan}(1985)}]{Padmanabhan}
\bibinfo{author}{\bibfnamefont{T.} \bibnamefont{Padmanabhan}},
  \bibinfo{journal}{Gen.\ Rel.\ Grav.} \textbf{\bibinfo{volume}{17}}
  (\bibinfo{year}{1985}) \bibinfo{pages}{215}.

\bitem[{\citenamefont{Sakharov}(1968)}]{Sakharov}
\bibinfo{author}{\bibfnamefont{A.~D.} \bibnamefont{Sakharov}},
  \bibinfo{journal}{Sov.\ Phys.\ Doklady} \textbf{\bibinfo{volume}{12}}
  (\bibinfo{year}{1968}) \bibinfo{pages}{1040}.

\bitem[{\citenamefont{Ashtekar, Rovelli and Smolin}(2000)}]{Ashtekar}
\bibinfo{author}{\bibfnamefont{A.}~\bibnamefont{Ashtekar}} \bibnamefont{and}
\bibinfo{author}{\bibfnamefont{C.}~\bibnamefont{Rovelli}} \bibnamefont{and}
  \bibinfo{author}{\bibfnamefont{L.}~\bibnamefont{Smolin}},
  \bibinfo{journal}{Phys.\ Rev.\ Let.} \textbf{\bibinfo{volume}{69}}
  (\bibinfo{year}{1992}) \bibinfo{pages}{237}.

\bitem[{\citenamefont{Smolin}(2004)}]{Smolin}
\bibinfo{author}{\bibfnamefont{L.} \bibnamefont{Smolin}},
  \bibinfo{journal}{Sci. Am.} \textbf{\bibinfo{volume}{290}}
  (\bibinfo{year}{2004}) \bibinfo{pages}{1}.

\bitem[{\citenamefont{Papapetrou}(1948)}]{Papapetrou}
\bibinfo{author}{\bibfnamefont{A.}~\bibnamefont{Papapetrou}},
  \bibinfo{journal}{Proc.\ Royal Irish Acad.} \textbf{\bibinfo{volume}{52A}}
  (\bibinfo{year}{1948}) \bibinfo{pages}{69}.

\endbib
%\end{thebibliography}

\end{document}